\documentclass[acmsmall,screen,10pt]{acmart}
\usepackage{booktabs}   
\usepackage{subcaption} 

\usepackage{cmap}
\usepackage[T1]{fontenc}

\usepackage[scaled=0.75]{beramono}
\usepackage{nanoml}
\usepackage{proof}
\usepackage{graphicx}
\usepackage{framed}
\usepackage{mdframed}
\usepackage{url}
\usepackage{wrapfig}
\usepackage{enumitem}
\usepackage{placeins}
\usepackage{caption}
\usepackage{makecell}

\newcommand{\keyterm}[1]{\textbf{#1}}
\newcommand{\newterm}[1]{\emph{#1}}
\newcommand{\packagename}[1]{\texttt{#1}}

\newcommand{\emptycxt}{\cdot}
\newcommand{\entails}{\vdash}

\newcommand{\sdef}{::=}
\newcommand{\sor}{\;|\;}

\newcommand{\isok}[1]{#1\: \text{okay}}

\newtheorem{hyp}{Hypothesis}

\lstdefinelanguage{JavaScript}{
  keywords={break, case, catch, continue, debugger, default, delete, do, else, false, finally, for, function, if, in, instanceof, new, null, return, switch, this, throw, true, try, typeof, var, void, while, with},
  morecomment=[l]{//},
  morecomment=[s]{/*}{*/},
  morestring=[b]',
  morestring=[b]",
  ndkeywords={class, export, boolean, throw, implements, import, this},
  sensitive=true
}

\lstnewenvironment{longcode}[1][]{%
  \lstset{language=nanoml,
             basicstyle=\scriptsize\sffamily,
             columns=flexible,
             breaklines=false
            floatplacement={tbp},captionpos=b,
           xleftmargin=12pt,xrightmargin=6pt,#1}}{}
\newcommand{\code}[1]{\mbox{\lstinline[language=nanoml,basicstyle=\scriptsize\sffamily,columns=flexible,mathescape=true]^#1^}}

\newcommand{\codebasic}[1]{\mbox{\lstinline[language=c,basicstyle=\scriptsize\sffamily,columns=flexible,mathescape=true]^#1^}}

\newcommand{\resetcaptionlengths}{
  \setlength{\abovecaptionskip}{3pt}
  \setlength{\belowcaptionskip}{1pt}
  \setlength{\floatsep}{1pt}
  \setlength{\textfloatsep}{2pt}
  \setlength{\intextsep}{1pt}}
\resetcaptionlengths

\setlist[itemize]{leftmargin=*}
\setlist[enumerate]{leftmargin=*}

\setcopyright{rightsretained} 
\acmPrice{}
\acmDOI{10.1145/3276492}
\acmYear{2018}
\copyrightyear{2018}
\acmJournal{PACMPL}
\acmVolume{2}
\acmNumber{OOPSLA}
\acmArticle{122}
\acmMonth{11}

\begin{document}

\title[One Tool, Many Languages]{One Tool, Many Languages: Language-Parametric Transformation with Incremental Parametric Syntax}


\author{James Koppel}
\affiliation{
  \institution{MIT}            
  \city{Cambridge}
  \state{MA}
  \country{USA}
}
\email{jkoppel@mit.edu}          

 \author{Varot Premtoon}
 
 \affiliation{
  \institution{MIT}            
  \city{Cambridge}
  \state{MA}
  \country{USA}
}
 \email{varot@mit.edu }

 \author{Armando Solar-Lezama}
 \affiliation{
  \institution{MIT}            
  \city{Cambridge}
  \state{MA}
  \country{USA}
}
 \email{asolar@csail.mit.edu}


\begin{abstract}

We present a new approach for building source-to-source
transformations that can run on multiple programming languages, based on a new way of representing programs
called \keyterm{incremental parametric syntax}. We implement this approach in Haskell in our \textsc{Cubix} system, and construct
incremental parametric syntaxes for C, Java, JavaScript, Lua, and
Python. We demonstrate a whole-program refactoring tool that runs on all of them, along with three smaller transformations
that each run on several. Our evaluation shows that (1) once a
transformation is written, little work is required to
configure it for a new language (2) transformations built this way
output readable code which preserve the structure of the original,
according to participants in our human study, and (3) our transformations can still handle language corner-cases, as validated on compiler test suites.
\end{abstract}



\maketitle


\section{Introduction}

\label{sec:has-dropbox-story}

In 2014, Dropbox had a massive refactoring to do. They wanted to let users log in with both a personal and corporate account on the same computer, but they had built the client assuming users only had one account. To change this, they needed to pass along information about which account each operation was for, and thread an \code{Account} parameter through tens of thousands of functions. Many program transformation experts could have readily built a tool for this, though it would have been a quite expensive task for one use-case, and Dropbox opted not to hire one. And so, in a company of over 100 engineers, the top project of the year was to tediously add parameters to functions.

Back in 2010, Facebook had a similar problem. All sorts of privacy
bugs were being exposed by the media, like weird combinations of
settings that would let someone view another user's private
photos. Facebook assembled a crack team; they needed this problem
fixed quickly, and made to never return. The privacy checks
were too haphazard: a bunch of conditionals every place where photos
may be displayed. They needed to move these all to one place:
private photos would never be fetched from the database in the first place. To do this, they needed to add a \code{ViewerContext} parameter to tens of thousands of functions. And so, for several weeks, every waking moment of several dozen of their top engineers was spent adding parameters to functions.

One might think that a clairvoyant entrepreneur in 2010 could have built a tool for this problem, and sold it to both Facebook and Dropbox. Alas, no, for Facebook's codebase was in PHP, while Dropbox's was in Python. And, with today's methods, building a similar program transformation tool for different languages requires {\bf building it separately for each language}.


We are not the first to notice the \emph{language-parametric
  transformation} problem of building a single transformation that can
run on multiple languages. Intuitively, this should be possible:
languages have a lot of similarities, and humans can readily apply the
same refactoring in many different languages. The challenge then is to
find some way to capture the similarities across languages, while
being flexible enough to express their differences.

The obvious approach is to convert
many languages into a single \emph{intermediate
  representation}. Unfortunately, doing so inevitably loses information.  While this
is fine for code-generation or analysis, it fails for source-to-source
transformations, which must produce an output
similar to the input. Instead, IR-based tools are known to "mutilate"
the program, such as by converting all for-loops into while-loops.

There is another line of work that promises the kind of flexible
representations needed: instead of building one representation to represent
all languages, having a different representation for each language,
but letting them share common fragments. This is the approach taken by
previous work on modular syntax
\cite{Bahr2011CompositionalDT,Zhang2015ScrapYB}, along with its cousin
work on modular interpreters \cite{liang1995monad} and modular
semantics \cite{Delaware2013ModularMM,Mosses2004ModularSO}. In
principle, these techniques could be used to do language-parametric
transformation, but the previous work does not scale to real
languages. All these approaches assume that
the entire language is built from these generic fragments. Hence, one
would have to do huge amounts of up-front work to define fragments
capable of representing all variations of each feature of modern
programming languages, and assemble them into representations for each language. The difficulty of developing language-parametric infrastructure has meant that previous work in this space, such as the work funded by the Dutch program on language-parametric program restructuring \cite{lammel2002towards,heering2004generic}, has all been for DSLs, toy languages, and language subsets.

This paper presents the first work that builds source-to-source transformations that run on multiple real languages. Our key insight is a new representation called \keyterm{incremental parametric syntax} (IPS). In incremental parametric syntax, languages are represented using a mixture of language-specific and generic parts. Like previous work on modular syntax, transformations deal only with the generic fragments. Unlike previous work, the implementer starts with a pre-existing normal syntax definition, and only does enough up-front work to redefine a small fraction of a language in terms of these generic fragments. Rather, they can \emph{incrementally} convert more of a language to generic fragments, as needed by new transformations. Best of all, since IPSs are defined as a "diff" to an existing syntax definition, implementations can re-use third-party language frontends.

We have implemented incremental parametric syntax in a Haskell framework called \textsc{Cubix}, and implemented support for 5 languages: C, Java, JavaScript, Lua, and Python. To evaluate \textsc{Cubix}, we built several program transformations that each run on multiple of those languages. We show transformations built in this style can have readable output, unlike IR-based approaches: our "Turing test" human study shows their output is no less readable than hand-transformed code. We show transformations built in this style can handle language corner-cases: the example transformations pass 100\% of compiler test suites, excluding some self-referential tests that should not pass ("assert function \code{foo} is declared on line 37") and tests that break the third-party parsers and pretty-printers. Finally, using \textsc{Cubix}, we created a prototype tool for threading variables throughout chains of function calls, as in Dropbox and Facebook's problem, and implemented it for all 5 language simultaneously (including Python, but not yet PHP).

\subsection{Why IRs Don't Solve Multi-Language Transformation}
\label{sec:why-irs-fail}

An old idea for building multi-language tools is to translate each language into some \emph{intermediate representation}. This works for writing analyses and code-generators, but is a poor fit for source-to-source transformation, which must preserve information. 

Conceptually, the IR approach to analysis is to provide a family of \code{lower} functions of type \code{C -> IR, Java -> IR}, etc, which transform each language into the IR, along with an \code{analyze} function of type \code{IR -> AnalysisResult}. Similarly, the IR approach to code generation provides a term of type \code{IR}, and \code{lift} functions of type \code{IR -> C, IR -> Java, IR -> Python}, etc. The natural extension to transformation is to implement a \code{transform} function of type \code{IR -> IR}, and compose it with the \code{lower} and \code{lift} functions to get language-specific transforms of type \code{Java -> Java, C -> C}, etc. But this makes a promise which is too good to be true: one can compose the \code{lower} and \code{lift} functions to get a translation from any language to any other!

The catch is that tools that implement this approach ``mutilate'' the program. Most commonly, the IR will be some kind of least common denominator of the supported languages, seen in frameworks like SAIL \cite{Dillig2009SailSA} and BAP \cite{Brumley2011BAPAB}, and bytecodes such as LLVM  \cite{Lattner2004LLVMAC} and the JVM \cite{lindholm2014java}. If the IR only supports \code{while}-loops, then any transformation through this \code{IR} will convert all loops into \code{while}-loops, even if the transformation has nothing to do with loops. Information about the original program has been lost. The alternative is for the IR to be a union of all concepts of the languages. The Clang AST, for instance, contains separate node types for both Objective-C and C++ exception-handling. This approach
essentially still requires the user to write a transformation separately for
each language: it can use the same node to represent similar constructs in different languages \textit{only if} they are exactly identical. And, among its many other drawbacks, it still loses information about what's \emph{not} in the program (e.g.: Java contains no pointer arithmetic, which simplifies analysis).

The end result is: because of these problems with conventional approaches, at time of writing, we are aware of no previous framework that allows the
user to define a single program transformation, run it on programs from multiple languages, and obtain output suitable for humans.

\subsection{Incremental Parametric Syntax}
\label{sec:intro-ips}

So, one-size-fits-all IRs don't work. Our solution is to find a way to apply parametric polymorphism to program transformations. The high-level idea of \emph{incremental parametric syntax} is to build transformations with the following functions:

\begin{longcode}[xleftmargin=55pt]
decompose_J$\;\;$::$\ $Java ->$\:\:$Generic $\,\,$\join$\;$ Remainder_J
decompose_C$\ \,$::$\ $C      ->$\:\:$Generic $\,\,$\join$\;$ Remainder_C
transform  $\;\,\,\,$::$\ $/\x. Generic \join x -> Generic \join x
recompose_J$\;\;\,$::$\ $Generic $\,\,$\join$\;$ Remainder_J$\,$ ->$\:$  Java
recompose_C$\ \,\,$::$\ $Generic $\,\,$\join$\;$ Remainder_C ->$\:$  C
\end{longcode}

Here, languages are decomposed into generic and language-specific parts. Then a transformation can be run on the generic parts, while preserving the rest of the program so that high-quality source code may be reconstructed. Unlike the common IR approach, these type signatures guarantee that a transformation cannot modify the language-specific parts, and the \code{decompose} and \code{recompose} functions cannot be used to translate one language into another.  And rather than construct the generic/language-specific decomposition up-front, IPS allows a programmer to begin with a third-party frontend for each language, and incrementally shift pieces of the language into the generic fragment as needed for new transformations. Hence, developers can add support for a new language in less than two days of work --- much of this time is spent looking at the language spec to understand how to model it in terms of generic components.

\label{sec:introduces-sort-injections}

The composition $X \bowtie Y$ is done using an approach known in term-rewriting as ``sum of signatures'' and
known in the functional-programming community as ``data types \`a la
carte'' \cite{swierstra2008data}. This approach can modularly define node types and mix-and-match them between languages, but does not let these nodes differ between languages: it cannot use the same notion of variable declarations to model both C declarations (which have types) and JavaScript ones (which do not). Similarly, in this approach, a generic assignment node cannot be used for both C/Java (where assignments are expressions) and in Lua/Python (where they are statements). We solve many problems with the new idea of \keyterm{sort injections}. Sort injections are deceptively simple: just add an \code{AssignIsExpression} node to C and an \code{AssignIsStatement} node to Python. Yet they complete sum-of-signatures by modularly specifying what \emph{edges} may be in an AST, and, in their general form, they solve many of the limitations of sum-of-signatures. Thanks to these sort injections, \textsc{Cubix} can take a pre-existing syntax definition for a language, and generate a new representation of the language which is fully isomorphic to the original, but replaces portions of the AST with generic nodes.

With each language expressed as an IPS, we can write a transformation parameterized on the nodes and sort injections it deals with. It can then be run on any language that has these nodes and sort injections, but will give a compiler error when used on one that does not. These transformations can be further parameterized on language-specific operations such as symbol resolution, allowing us to build sophisticated multi-language transformations that can still handle many language-specific corner-cases.


\subsection{Contributions}

Overall, our paper introduces the following new ideas:

\begin{itemize}
\item We present the concept of \keyterm{parametric syntax}, which allows the user to define a family of representations
  specific to different languages, but source-to-source transformations that can run on many of them. We further present techniques for
  \keyterm{incremental parametric syntax}, which allows the user to
  achieve this with little work, given 3rd-party parsers and pretty-printers.
\item We develop the concept of \keyterm{sort injections}, which modularly specify which edges may be in an AST, and hence provide a
  typed and modular way to intermix language-specific and generic
  fragments.
\item We present an algorithm for automatically converting a data type into the sum-of-products representation and its implementation in the \packagename{comptrans} code-generation library.
\end{itemize}

\noindent We use these ideas to produce the
following results.

\begin{itemize}
\item We demonstrate using incremental parametric syntax
  to define a representation for C, Java, Python, JavaScript, and Lua.
  We show how we can define transformations that can run on all of them, including a realistic whole-program refactoring tool and complicated transformations based on control-flow, yet still achieve results that are as readable as hand transformed code.
\item  We present the results of a human study comparing the output of our transformations against hand-transformed code. These were identical 20\% of the time, and, of the rest, judges gave ours higher average scores for readability.
\item We present the RWUS (Real World, Unchanged Semantics) suite, consisting of 50 programs across
  5 languages randomly drawn from top GitHub projects, together with
  test suites thorough enough to detect almost any modification that changes
  program semantics.
\item We present the IPT (``Interprocedural Plumbing Transformation'')
  tool, a whole-program refactoring for threading variables through
  chains of function calls, as in the Dropbox and Facebook
  stories. We show how we used \textsc{Cubix} to simultaneously build this tool for 5 languages.

\end{itemize}

\subsection{Limitations}

\label{sec:limitations}

This paper is the first to present language-parametric transformations that work on multiple real languages, greatly surpassing previous attempts which were built for DSLs and language subsets such as \citet{lammel2002towards}. Nonetheless, we have not solved all problems relating to language-parametric tools. Here are several non-goals of this paper:

First, we do not address usability. \textsc{Cubix} is not a tool for the lay-programmer. New undergrads on the project take about two months part-time to learn enough generic programming to begin using the system. Furthermore, we emphasize that this paper addresses only the "1-to-n" problem of extending a transformation to many languages. It is already very hard to bring a transformation to $100\%$ correctness for one language, let alone $5$; this is why there are only $4$ transformations in this paper.

Second, this paper focuses on techniques for transformation, not
analysis or specific transformations. The contribution of this paper
is the \textsc{Cubix} framework and the techniques used to make
language-parametric transformation possible, not the example
transformations in this paper. These transformations take the results
of program analyses as input, and so our work dovetails with
techniques for multi-language analysis, but we do not address analysis
in this paper. Note also that it is easy to integrate a language-parametric transformation with multiple language-specific analyses, so long as they provide a common interface.

Third, we have done no performance-engineering. The current implementation has substantial overhead, though we have lots of ideas for optimization, and it's still fast enough to transform and run all $2782$ JavaScript tests in 5 minutes on the first author's laptop.

Building language-parametric tools is a long-term goal. There is still work to be done on verifying language-parametric transformations, using work on modular type systems to create type-aware transformations, dealing with differing memory models, preservation of formatting, etc.

\section{Overview}

In this section, we show how our approach allows constructing language-parametric transformations, and the work required to add support
for a new language. In Section \ref{sec:building-hoisting}, we explain the construction of a
transformation called ``declaration hoisting,'' and how it is
configured to run on several languages. Section
\ref{sec:modularizing-c} then explains how to create an incremental
parametric syntax for C. In the language of Section \ref{sec:intro-ips}, Section \ref{sec:building-hoisting} defines \code{transform}, while Section \ref{sec:modularizing-c} defines \code{Remainder_C}, \code{decompose_C}, and \code{recompose_C}.

\vspace{0.1em} 

\subsection{An Elementary Hoisting Transformation}

\label{sec:building-hoisting}

In this section, we describe the construction of a simplified transformation for
\newterm{declaration hoisting}, and how with a small amount of configuration, we can apply it to C, Java, and JavaScript. This transformation showcases the versatility of our approach: although it totals only 22 lines for the transformation plus 30 lines for the language-specific code, it handles multiple language corner-cases, and achieves a high pass rate on the compiler validation test suites. Full code for the general portion is given in Section \ref{sec:impl-elem-hoist}.

The declaration hoisting transformation moves all variable
declarations to the top of the scope, using normal assignments to
initialize them. The end result is similar to how C89
requires programs to be written. Figure
\ref{fig:hoisting-example} gives an example C program and its hoisted version. The elementary hoisting transformation of this section is a simplified version of the hoisting transformation in our benchmarks (\ref{sec:benchmark-transformations}), which also supports Lua and handles shadowing. Neither supports Python because Python lacks variable declarations.

\begin{figure}
\begin{center}
\lstset{
  numbers=left,
  numbersep=5pt, 
  basicstyle=\scriptsize\sffamily,
  columns=flexible,
  breaklines=false
  floatplacement={tbp},captionpos=b,
  xleftmargin=8pt,xrightmargin=8pt
}
\begin{tabular}[t]{|p{0.3\textwidth}|p{0.3\textwidth}|}
\hline
\begin{lstlisting}[language=c]
int f(int a,int b,int s) {
  int t1 = 0, t2 = 1;
  if (s) {
    int r1 = t1*a+t2*b;
    return r1;
  }
  int r2 = t2*a+t1*b;
  return r2;
}
\end{lstlisting}
 &
\begin{lstlisting}[language=c]
int f(int a,int b,int s) {
  int t1, t2; int r2;
  t1 = 0; t2 = 1;
  if (s) {
    int r1;
    r1 = t1*a+t2*b;
    return r1;
  }
  r2 = t2*a+t1*b;
  return r2;
}
\end{lstlisting} \\
\hline
\end{tabular}
\caption{An example of hoisting a C program}
\label{fig:hoisting-example}
\end{center}
\end{figure}

\paragraph{Setting the syntactic constraints}

The user first writes a type signature declaring the general syntactic constructs a language must have to use this transformation: variable declarations, assignments, blocks, and identifiers. The type signature also requires that assignments and variable declarations must be valid members of blocks --- these are {\it sort injections}, as described in Section \ref{sec:modularizing-c}. 
Figure \ref{fig:hoist-constraints} in Section \ref{sec:implementation} gives the code listing these constraints.

\paragraph{Language-specific operations}

Variable initializations and assignment RHSs can be different. A Java array initialization \codebasic{int[] x = \{1,2,3\};} becomes \code{x = new int[]\{1,2,3\};}. C variable declarators have different abstract syntax from C lvalues. To deal with these, the transformation takes as a parameter two language-specific operations, \code{varInitToRhs} and \code{varDeclBinderToLhs}.

\paragraph{Writing the transformation}

The transformation traverses every block in the program. At each block, it checks if each block item is a variable declaration. If so, it splits the declaration into one
without initialization, and into a sequence of zero or more
assignments. The extracted assignments are
inserted where the variable declarations lay previously, while the extracted variable declarations are prepended to the front of the block. Figure \ref{fig:elem-hoist-code} in Section \ref{sec:implementation} gives this code verbatim.

\paragraph{Dealing with language subtleties}

The hoisting transformation deals with several subtleties through the language-specific operations, but we give another one here: In JavaScript directives such as \codebasic{"use strict";} must be placed at the top of a block to have effect; hoisting something above it can break the
code. Perusing the spec, we saw directives are essentially treated as a separate kind of syntax, so we modified the representation of JavaScript to store them
separately. This fixed bugs in multiple transformations. More examples
are given in the figures in Section \ref{sec:benchmark-transformations}.

While simple, the elementary hoisting transformation in this section runs on three languages, deals with multiple language subtleties, and has a $98.4\%$ pass rate on compiler test suites (compared to the $100\%$ pass rate of the real version). Overall, these techniques allow transformations for different languages to share code to the extent that the two languages are syntactically similar. The rest of this paper gives more interesting transformations that also make use of static analysis and control-flow information.

\subsection{Modularizing C}

\label{sec:modularizing-c}

In Section \ref{sec:building-hoisting}, we outlined how to build a hoisting transformation which works on any language that
contains some common notion of variable declarations, assignments, and blocks. We now show how to construct an incremental parametric syntax for C, in which parts
of the language are recast in terms of these common components.

\begin{figure*}
\begin{center}
\includegraphics[scale=0.13]{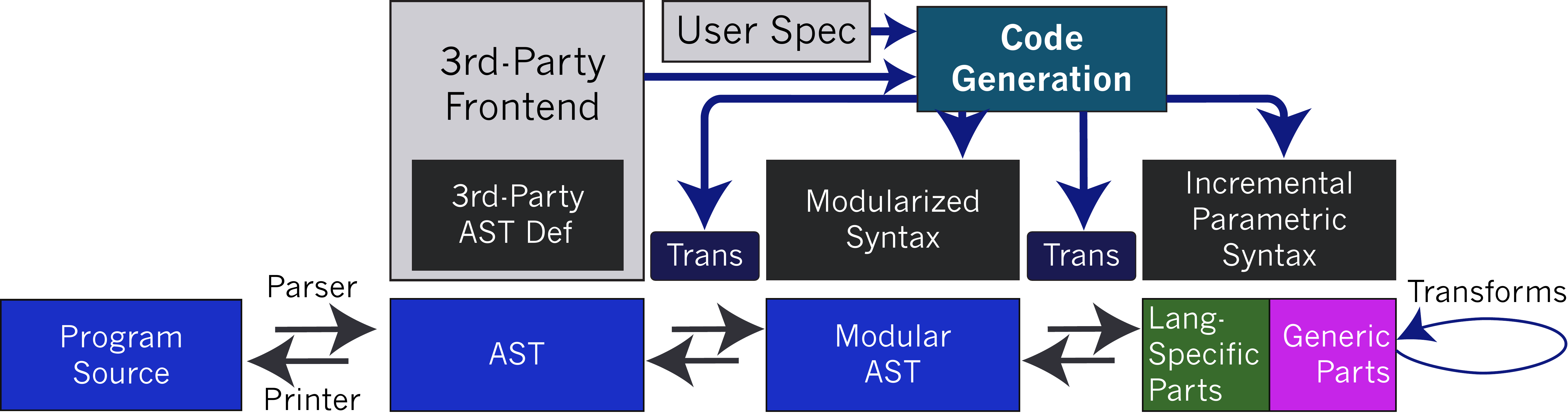}
\end{center}
\caption{Architecture of Cubix}
\label{fig:architecture-diagram}
\end{figure*}

Our approach gives a language three representations. The starting point is some already-existing syntax definition of the language from a 3rd-party library, with its
accompanying parser and pretty-printer. For C, we use Haskell's
\packagename{language-c} library \cite{authors2016language-c-0.5.1}, which defines C's abstract syntax as a set of mutually recursive algebraic data types like \code{CExpression} and \code{CAssemblyStatement}. Next is the ``modularized'' representation, which gives the exact same set of data types, but as independent signatures that do not reference each other. The sum of these signatures is isomorphic to the \packagename{language-c} abstract syntax definition. This make it easy to sum together a different set of signatures, replacing some of the C-specific data types with generic ones, yielding the third representation, the incremental parametric syntax. These three representations are mutually isomorphic, and translations between them are derived mostly automatically: the user only writes code to translate between removed node types and their generic equivalents. Figure \ref{fig:architecture-diagram} depicts how the representations and translations are generated, and how a program is transformed
through each of them at runtime.

The remainder of this section explains how to construct an incremental parametric syntax for C.

\paragraph{Modularized representation}

For each algebraic data type in the C abstract syntax, the user must
generate a new data type representing nodes of that sort inside an
arbitrary AST (a \keyterm{signature} for that node). Combining these
give a new representation identical to the original, but made of
independent components. The user generates these definitions
completely automatically, using the Template Haskell code-generation
engine. Section \ref{sec:modular-syntax} explains how we represent and combine signatures, while Section \ref{sec:comptrans} explains the data type transformation in more detail.

\paragraph{Incremental parametric syntax: Nodes}

The hoisting transformation is built on general components for
assignments, variable declarations, and blocks. The user will need to
replace these components of C, but no others, with their corresponding
generic components, yielding the incremental parametric syntax. This
is incremental because the user will revisit this step as more
components of C need to be genericized to support new transformations.

To genericize these components, the user first compares their
definitions in the C specification to the specification of the generic
components, making sure the latter can model the former. To customize
the generic \code{VarDecl} node to C, the user must create a new node
of sort \code{VarDeclAttrsL} containing the C-specific components of a
variable declaration (type and storage specifiers, assembly name, and
attributes). The user does similarly for a couple other C constructs.

\paragraph{IPS: Sort injections}

The user now finishes customizing the generic components to C by
specifying where they fit in the C syntax. The user indicates that
generic assignments may be used as C expressions, while C expressions
form the RHS of assignments. The user does this by e.g.: generating a
\code{AssignIsCExpr} node. This establishes an injection from terms of
sort Assign to terms of sort CExpr, which we call a \keyterm{sort
  injection}. The user does similar to place the other generic nodes
within the C syntax. \textsc{Cubix} generates nodes witnessing these
sort injections


\paragraph{IPS: Putting it together}

\begin{figure}
\begin{center}
\includegraphics[scale=0.20]{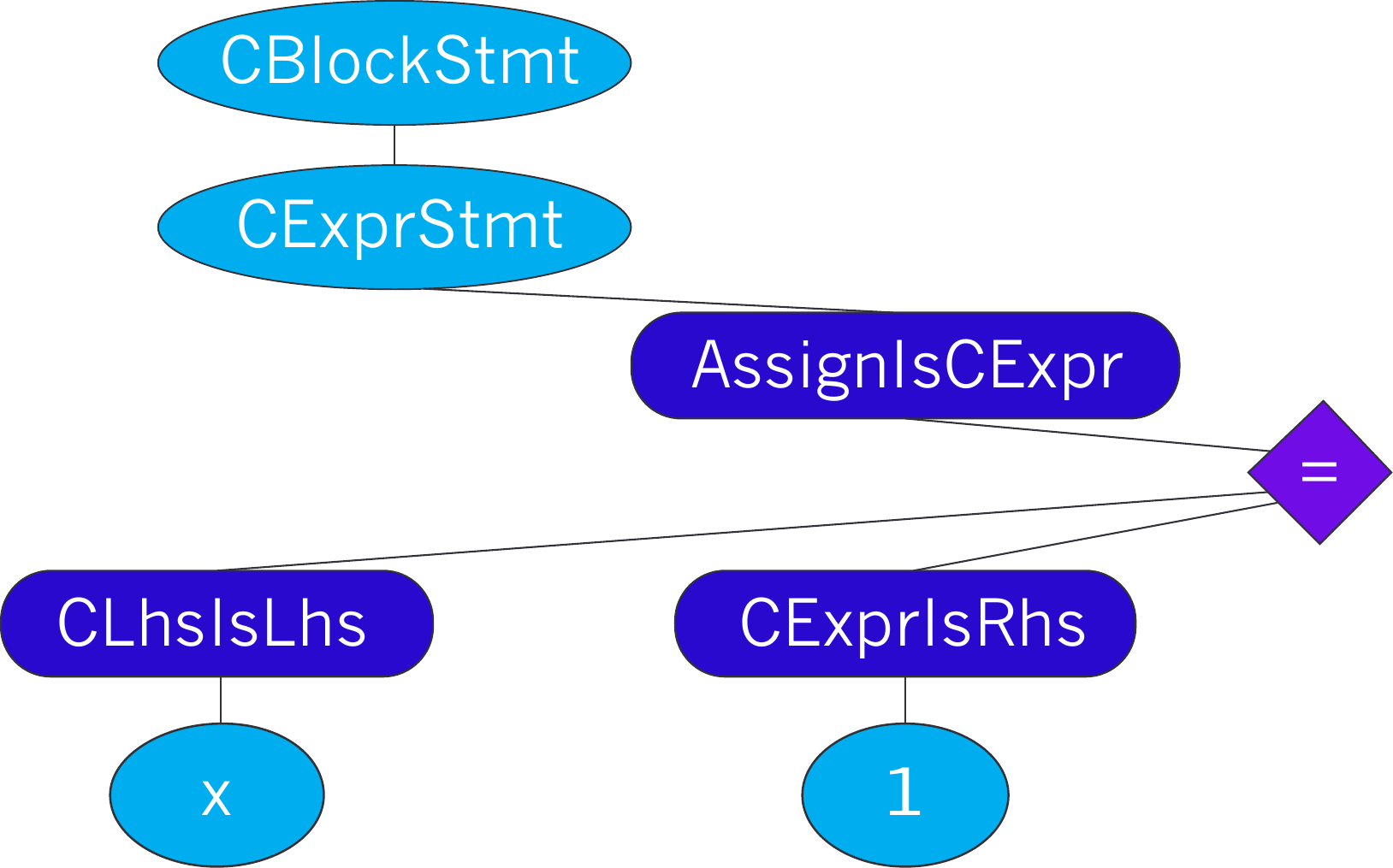}
\end{center}
\caption{A term in the incremental parametric syntax for C. The ellipses (light blue) represent language-specific nodes; rhombi (purple) represent generic nodes; rounded rectangles (dark blue) represent sort injection nodes.}
\label{fig:modular-tree}
\end{figure}

The user now defines the incremental parametric syntax for C by writing a couple lines of Template Haskell that takes the list of signatures in the modularized syntax, subtracts the replaced nodes, and adds the generic components and sort injection nodes. This code is given in Figure \ref{fig:ips-gen-code} in Section \ref{sec:impl-lang}. The sum of these signatures is the signature for the IPS for C, and the terms of this signature are given by its type-level fixpoint. These terms resemble Figure \ref{fig:modular-tree}, showing the mixture of C-specific and generic nodes, with sort injections between them.


\paragraph{IPS: Translations}

The user writes instances of the \code{trans} and \code{untrans} operators between the nodes that have been removed from C, and the generic ones that replaced them.  Generic programming deals with the nodes shared between the IPS and the modularized syntax, giving translation functions between the two representations. Our actual implementation of these translations for C totals 130 lines of Haskell code, about 40 of which are boilerplate.

\setlength{\belowcaptionskip}{8pt}
\resetcaptionlengths

\section{Core Ideas}

In this section, we explain the core new ideas that make our
language-parametric transformations possible. Section \ref{sec:modular-syntax} gives background on modular syntax, used in the rest of this section Section \ref{sec:ips} presents the terminology and goals of incremental parametric syntax. We achieve this through the concepts in the following sections: Section \ref{sec:sort-injections} presents sort injections, and Section \ref{sec:comptrans} explains the translation of a syntax into its modularized version.

\subsection{Background: Data Types \`a la Carte}
\label{sec:modular-syntax}

\begin{figure}
\begin{center}
\includegraphics[scale=0.25]{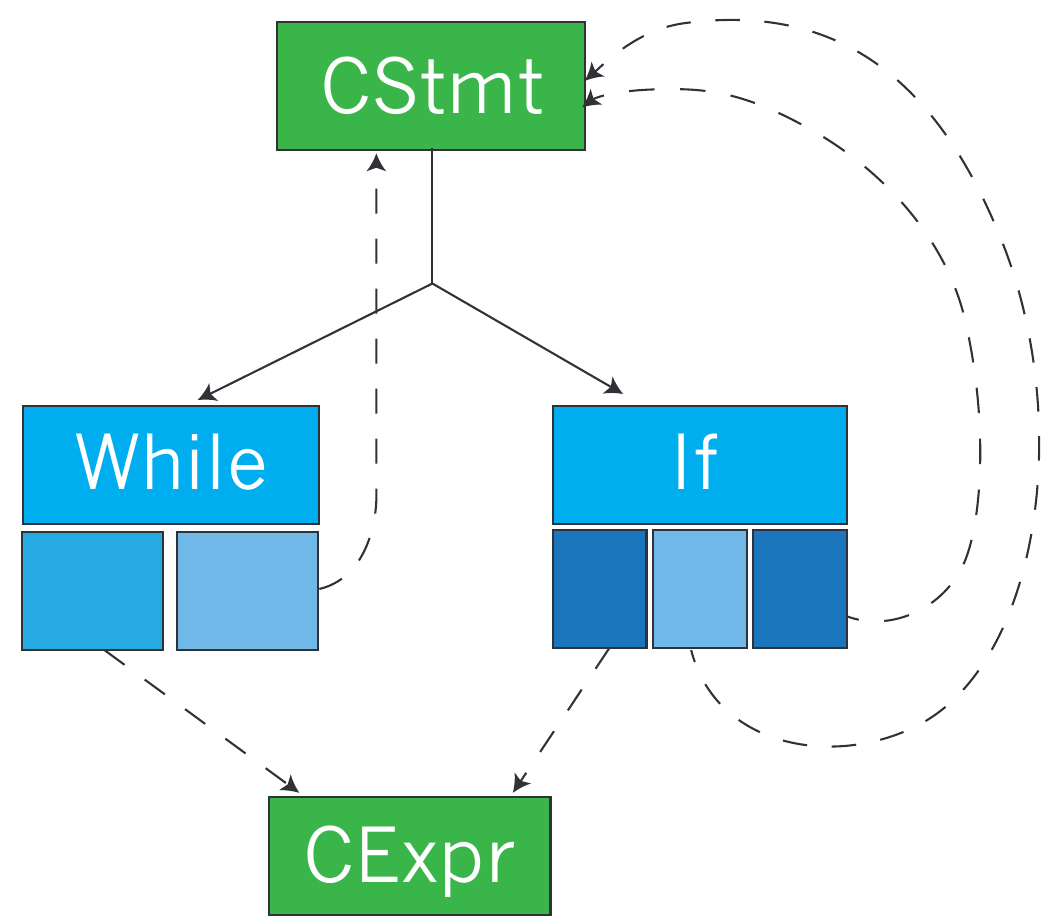}
\end{center}
\caption{Fragment of a typical representation of C. The solid arrows represent the instance-of relationship; dotted represent containment.}
\label{fig:c-repr-tangled}
\end{figure}

The basic idea of modular syntax is simple: languages should be defined by a set of nodes, and the same node can appear in many languages. So, a transformation to swap the two branches of an if-statement should be runnable on any language with if-statements.

Unfortunately, in common representations of syntax, whether as an algebraic data type (ADT) like the fragment in Figure \ref{fig:c-repr-tangled}, or as a set of classes, this is not possible. The problem is mutual recursion between types. A C if-statement contains C expressions, which can contain C statements. So, the node for C if-statements is tied to definitions for all other C statements. The structure of code follows the structure of data, and so a traversal written over this type will also be coupled to all C statements.

Even without the mutual recursion, trouble arises as soon as one uses
a fixed type like $C \rightarrow C$ or $\text{Java} \rightarrow \text{Java}$ for a program transformation. The reason goes back to the basic theory of subtyping. Producer functions of type $A \rightarrow C$ are \emph{covariant} in $C$, meaning new cases can be added to $C$ without changing the function. Consumer functions of type $C \rightarrow A$ are \emph{contravariant} in $C$, meaning cases can be removed from $C$ without changing the function. But functions of type $C \rightarrow C$ are \emph{invariant}, meaning the code will break if any cases are added or removed from the language. Techniques such as the visitor pattern can help, but introduce new limitations (discussed by \citet{lammel2003}), and do not allow for a multi-language transformation so long as the types are tied together. Switching to a dynamically-typed language also does not help; removing the types does not remove constraints over the data.

What does help is removing the recursion from the syntax definitions, and using parametric polymorphism for the types. Mathematically, an ADT is defined in three stages: first data is tupled into a constructor; then many constructors are summed into a signature, a list of node types with unspecified children; and then a fixpoint is taken over the signature, yielding recursive trees. The idea of the sum-of-signatures representation, known in the functional programming community as data types \`a la carte (DLC) \cite{swierstra2008data}, is to defer the fixpoint operation. The programmer instead programs against signatures, which are not recursive, and can be modularly combined.

\begin{figure}
\begin{center}
\includegraphics[scale=0.13]{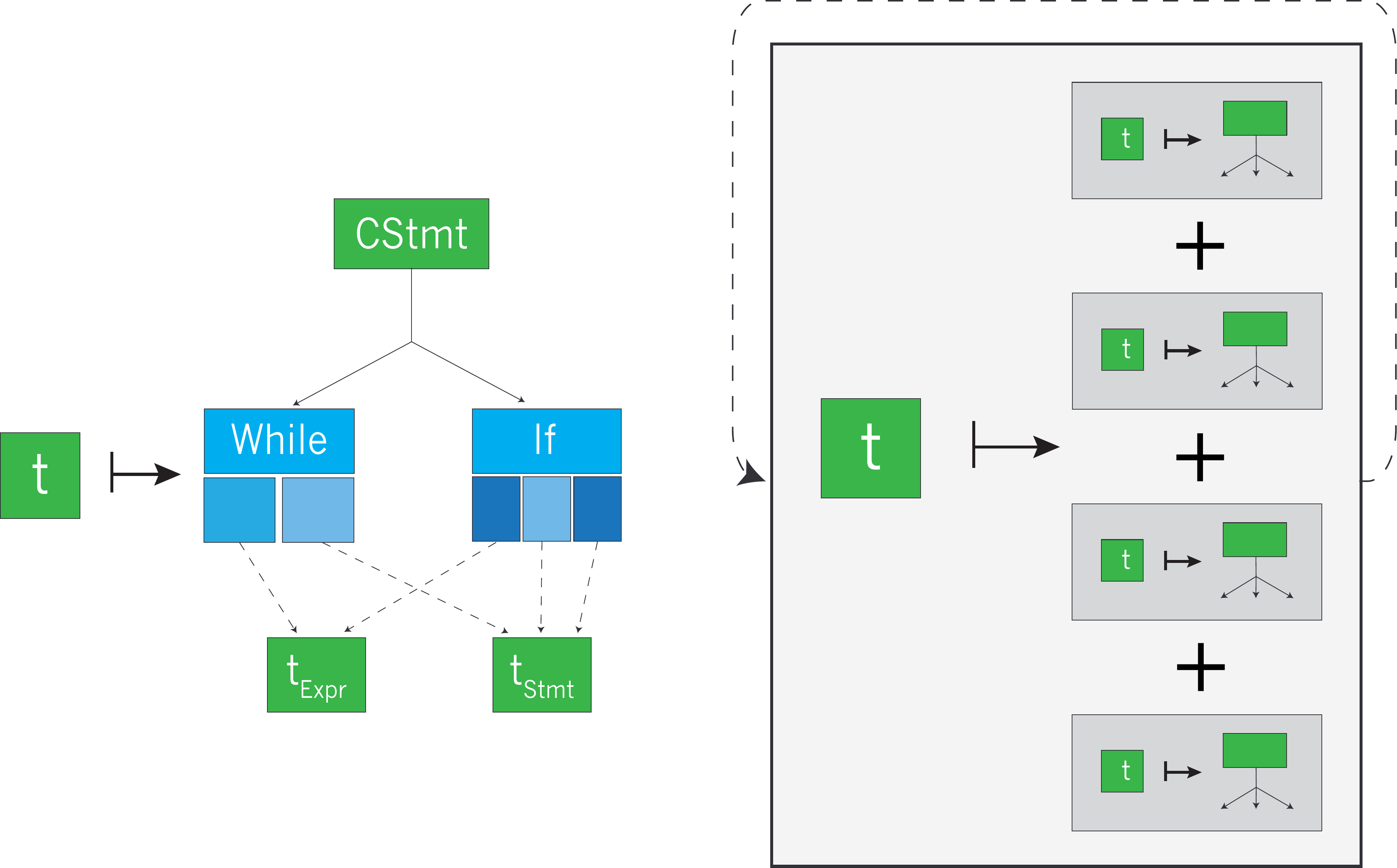}
\end{center}
\caption{In DLC, a language
is represented by a list of subsignatures like the one on the left. Each signature has a type variable for subterms, in lieu of self-reference. The subsignatures are combined into
a signature for the whole language, which is then closed by specifying that allowed subterms of terms of this signature are other terms of this signature (right).}
\label{fig:c-repr-para}
\end{figure}

In DLC, a signature takes the form of a data type similar to conventional abstract syntax, but where all recursive terms have been replaced with a type variable, so that the type of children may be specified later. Each signature may represent an independent fragment of a language; these signatures may be freely summed into a signature for an entire language, and then closed recursively, as depicted in Figure \ref{fig:c-repr-para}. Figure \ref{fig:example-dlc} shows an example of a term written in DLC, taken from \citet{swierstra2008data}.

\begin{figure}
\begin{center}
\begin{framed}
\begin{longcode}[numbers=left]
data Add e = Add e e
data Val e = Val Int
data (f :+: g) e = Inl (f e) | Inr (g e)
data Term f = Term (f (Term f))

type ExpSig = Add :+: Const
type Exp = Term ExpSig

addExample :: Exp
addExample = Term (Inl (Add (Term (Inr (Val 118))) (Term (Inr (Val 1219)))))
\end{longcode}
\end{framed}
\caption{Using data types \`a la carte to present the expression \code{118+1219}, with addition and constant nodes defined in separate fragments.}
\label{fig:example-dlc}
\end{center}
\end{figure}

Data types \`a la carte generalizes easily to multiple sorts: have a type variable for terms of sort \code{Stmt}, another for terms of sort \code{Exp}, etc. This unfortunately makes it difficult to add new sorts, or to have languages with different numbers of sorts. The insight of  \citet{yakushev2009generic} is to merge these into a single higher-order type variable $t$. Subscripting $t$ with various labels gives the type of terms of a certain sort: $t_{\text{\code{Stmt}}}$ represents terms of sort \code{Stmt}, $t_{\text{\code{Exp}}}$ represents terms of sort \code{Exp}, etc. But \code{t} itself is a single variable, representing terms of all sorts. Figure \ref{fig:modularized-example} shows signatures following this pattern.

\setlength{\belowcaptionskip}{4pt}
\resetcaptionlengths

\subsection{Incremental Parametric Syntax}

\label{sec:ips}

As explained above, functions of type $C \rightarrow C$ have a type which is {\it invariant} in $C$. That is, in general, the code for any function that consumes and produces a value of type $C$ will break when the definition of $C$ is modified. So, instead of using a fixed type, the way to write a function that can transform many data types is with parametric polymorphism. For instance, the sort function of type $\forall x. \text{Ord x} \Rightarrow  [x] \rightarrow [x]$ works over lists of any data type that supports comparison, and, after inlining, is just as efficient as a sort function written for each data type. Our goal is to bring this combination of generality and specialization to program transformation.

Let $F_1, \dots, F_n$ be fragments that may be contained in many languages. We define a \keyterm{parametric syntax} $\mathcal{S}$ for a language as any representation that supports an operation $\prec$ such that a transformation over any language containing $\overline{F_i}$ may be written  $\forall x. \overline{F_i \prec x} \Rightarrow x \rightarrow x$. This gives a name to previous work: any language written in DLC is a parametric syntax.

But the drawback of previous incarnations DLC and other forms of modular syntax is that language definitions in those styles are what we term a \keyterm{fully parametric syntax}, meaning that the syntax must be written entirely in terms of generic fragments.

More formally, a fully parametric syntax is any representation satisfying:

\begin{enumerate}
\item There is some combination operator $\bowtie$ which merges fragments. The $\bowtie$ operation must satisfy the property: if $F \prec G$ or $F \prec H$, then $F \prec (G \bowtie H)$ .
\item Each syntax definition is built entirely by combining generic fragments. That is, $\mathcal{S}$ is a fully-parametric syntax if it can be written $\mathcal{S} \doteq G_1\bowtie \dots \bowtie G_m$, where each $G_i \in \{F_1, \dots, F_n\}$. 
\end{enumerate}

Defining a fully parametric syntax for a language requires a large amount of up-front labor. Incremental parametric syntax lowers this initial barrier.

We say that $\mathcal{S}$ is an \keyterm{incremental parametric syntax} if there is a non-parametric syntax $\mathcal{T}$
and a ``fragment removal'' operator $\setminus$ such that $\mathcal{S}$ may be expressed: \[
  \mathcal{S} \doteq (\mathcal{T} \setminus F_1 \setminus \dots
  \setminus F_m) \bowtie G_1 \bowtie \dots \bowtie G_n \]

An incremental parametric syntax allows the user to start with a pre-existing syntax definition, replace some components with their generic equivalents, and then write transformations against the generic components. Given the complexity of a production language, this approach is necessary for getting a language-parametric transformation running on real languages in a reasonable amount of time.

In our instantiation of incremental parametric syntax, we use the signature subsumption and sums from data types \`a la carte to provide the fragment subsumption ($\prec$) and $\bowtie$ operators. We use new ideas for the $\setminus$ operator: convert the existing syntax into a sum of language-specific signatures (Section \ref{sec:comptrans}), and then use signature subtraction. Additionally, to add generic fragments, one must also add new nodes to reshape the grammar to accept them (Section \ref{sec:sort-injections}),

Parametric syntax closely relates to the Expression
Problem~\cite{wadler1998expression}, which concerns being able to separately extend a language with new terms and new operations. Any incremental parametric syntax is also a solution to the Expression Problem, as it allows a language to be extended with new terms and operations. However, a solution to the Expression Problem need not allow for expressing multiple languages. As parametric syntax is our name for a family of existing approaches, discussion of how parametric syntax solves the Expression Problem can be found in the DLC paper \citep{swierstra2008data}.

\subsection{Sort Injections}
\label{sec:sort-injections}

Using data types \`a la carte, we can modularly specify which nodes may
be in a language, and replace them with generic ones. However, similar
nodes in different languages may interact differently with the
rest of the language. Assignments are expressions in C/Java and
statements in Lua/Python. Most languages have various assignment
operators like \code{+=}, but Lua does not. Rarely will a generic node be an exact fit for a construct already in
a language. Instead, it must be customized for that language.

We solve this with sort injections. A \keyterm{sort injection} from A
to B is an injective function from terms of sort A to terms of sort B,
together with its partial inverse. C and Java have a sort injection
from \code{Assign} to \code{CExpr} and \code{JavaExpr} respectively,
while Lua and Python have ones from \code{Assign} to their respective
statement sorts. And all languages except Lua have a sort-injection
from their respective language-specific assignment-operation sorts to
the generic assignment-operation sort.

The most straightforward way to provide such a sort injection is via a \keyterm{sort injection node}, an unary production of sort B with a single child of sort A. Figure \ref{fig:example-sort-inj} gives an example sort injection and node from generic assignments to C expressions. 

So, while DLC modularizes which nodes may be in a languages,
sort-injections modularize the edges. Adding the \code{AssignIsCExpr}
node from Figure \ref{fig:example-sort-inj} to a syntax definition is equivalent to allowing a parent-child edge from anything that contains C expressions to assignments.

\begin{figure}
\begin{center}
\vspace{2pt}
\includegraphics[scale=0.20]{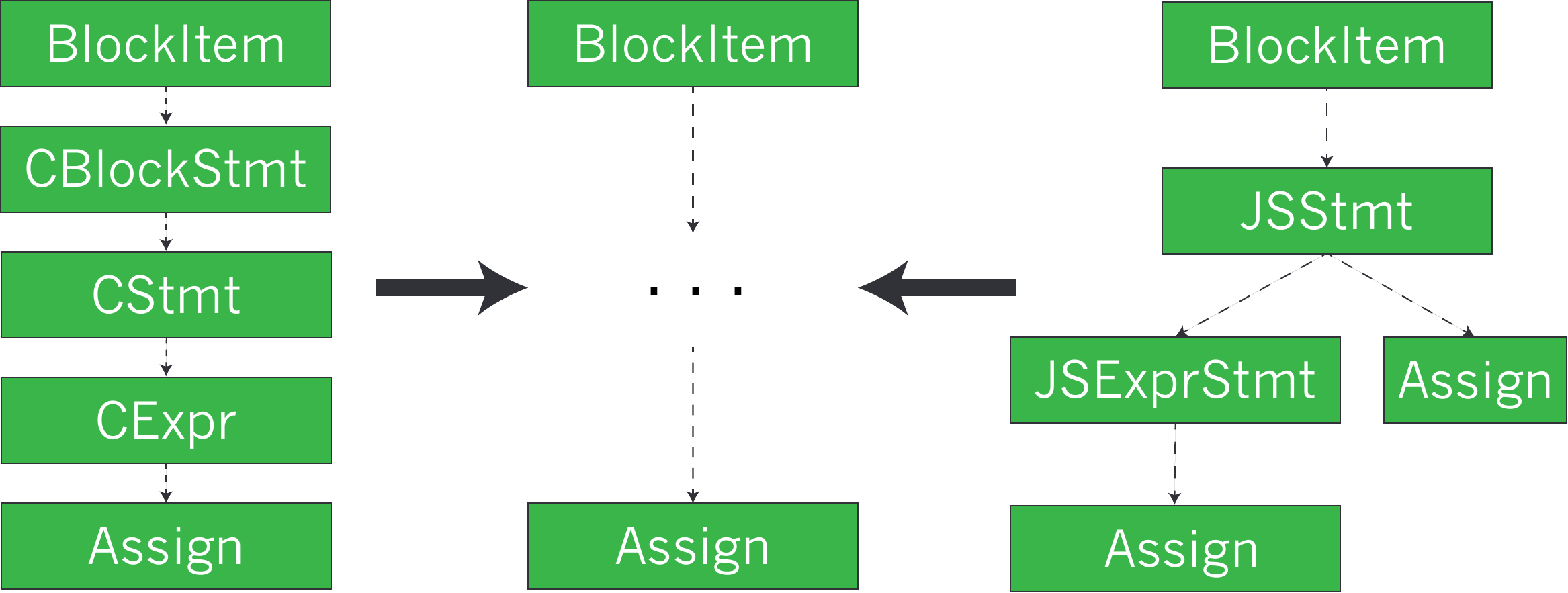}
\caption{Sort injections from \code{Assign} to \code{BlockItem}}
\label{fig:abstracting-syntax}
\end{center}
\end{figure}

Sort injections also serve an additional purpose: abstracting over
intermediate nodes. In all supported languages, assignments may be
used as top-level items in blocks. However, this occurs through a
chain of language-specific nodes. Block statements in C may not be
assignments directly, but they can be ordinary statements, which may
be expression statements, which may be assignments. And, in the 3rd-party
JavaScript frontend used by \textsc{Cubix}, there are actually two
(semantically-equivalent) ways that assignments may be statements. All
this can be abstracted into the constraint: there is a sort
injection from assignments to block items. Figure
\ref{fig:abstracting-syntax} illustrates this example.

\begin{figure}
\begin{center}
\vspace{2pt}
\includegraphics[scale=0.20]{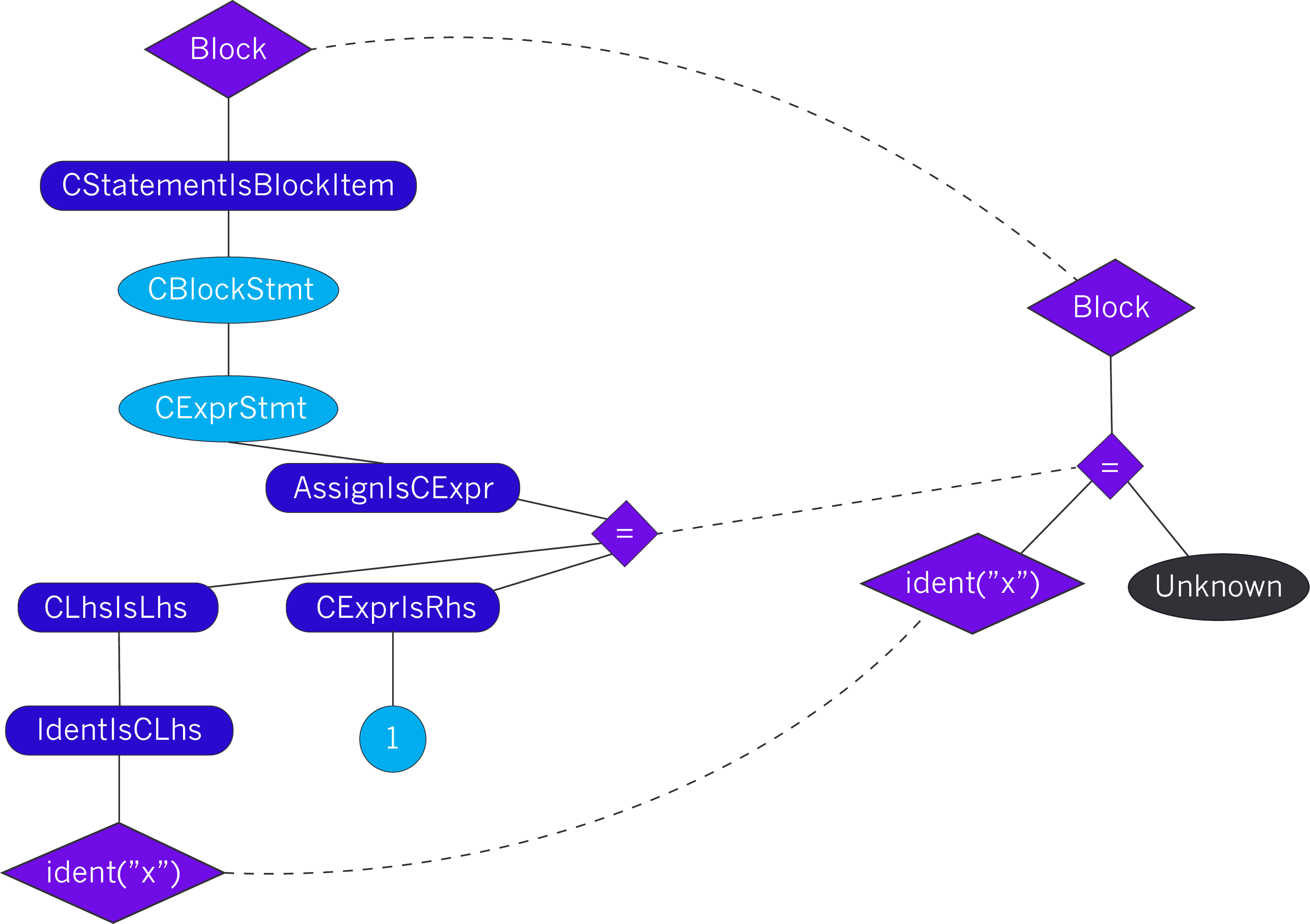}
\caption{Blowing down a tree}
\label{fig:abstracting-tree}
\end{center}
\end{figure}

In a transformation such as Hoist that works on languages with a sort
injection from assignments to block items, the transformation has the
ability to place assignments into blocks, and to check if a block item
is an assignment. So one can think of this transformation as working
not on the original tree but on a ``blown-down'' tree, which only
contains these generic nodes. Figure \ref{fig:abstracting-tree}
shows an example of a blown-down tree. This is similar to the theory views seen in Maude \citep{clavel2002maude}, and to the
homeomorphic embedding in term rewriting \citep{baader1999term}.

\begin{figure}
\begin{center}
\begin{framed}
\begin{longcode}[numbers=left]
data AssignIsCExpr t l where
  AssignIsCExpr :: t AssignL -> AssignIsCExpr t CExprL
instance (AssignIsCExpr :<: f) => InjF (Term f) AssignL CExprL where
  injF = AssignIsCExpr
  projF x = case project x of
    Just (AssignIsCExpr x) -> Just x
    _               $\;$       -> Nothing
\end{longcode}
\end{framed}
\caption{Sort injection node and its associated sort injection}
\label{fig:example-sort-inj}
\end{center}
\end{figure}

\subsection{Modularizing a Syntax Definition}

\label{sec:comptrans}

The preceding sections gave some of the techniques of incremental parametric syntax; we now show how to convert an existing syntax definition so that it can be incrementally generalized. The key idea is to transform an existing syntax definition $\mathcal{T}$ into the combination $F_1 \bowtie \dots \bowtie F_m$. This provides the final component of incremental parametric syntax, as $\mathcal{T} \setminus F_i$  can be defined by simply removing $F_i$ from the combination. 

The ADT modularization transformation is most easily explained by an example: it transforms the ADTs on the left side of Figure \ref{fig:modularized-example} into the generalized algebraic data types (GADTs) on the right. The GADTs stand independently, with no recursion between them. Instead of a recursive reference to the \code{Arith} type, for example, the type \code{t ArithL} can be read "Terms of sort \code{ArithL}, which will be specified later." But when those terms are specified, and the independent types are combined back together, the result type \code{Term $\;$(Arith :+: Atom :+: Lit) ArithL} is isomorphic to \code{Arith}. Figure \ref{fig:fixed-term} shows how these types are combined.

In our instantiation of incremental parametric syntax, this means converting a syntax definition into DLC. For syntax definitions given as mutually recursive algebraic data types, this is quite easy to do. The recursive knot is already tied in a separate step in the metatheory; the ADT modularization transformation just puts that in code as well. 

The transformation generalizes easily from this example. We give a
full formal definition in Appendix \ref{app:adt-modularization}
 and implement it in our \packagename{comptrans} tool.

\begin{figure}
\begin{center}
\begin{tabular}[t]{|p{0.30\textwidth}|p{0.35\textwidth}|}
\hline
{\begin{longcode}[numbers=left]
data Arith = Add Atom Atom

data Atom  =  Var String
           | Const Lit
           | Parens Arith
           
data Lit = Lit Int
\end{longcode}}
 & 
{\begin{longcode}[numbers=left]
data ArithL; data AtomL; data LitL
data Arith t l where
  Add :: t AtomL ->$\;$ t AtomL
                  -> Arith t ArithL
data Atom t l where
  Var    :: String   -> Atom t AtomL
  Const  :: t LitL   -> Atom t AtomL
  Parens :: t ArithL -> Atom t AtomL
data Lit (t :: * -> *) l where
  Lit :: Int -> Lit t LitL
\end{longcode}} \\
\hline
\end{tabular}

\caption{Example input (left) and output (right) of \code{comptrans}.}
\label{fig:modularized-example}
\end{center}
\end{figure}

The modularized representation has several other benefits, even when writing transformations for only one language. For instance, it allows giving a
type \code{Term Sig l -> Term Sig l} to sort-preserving rewrites that can be applied to terms of any
sort, and also enables many generic-programming techniques. See \citet{Bahr2011CompositionalDT} for a
full discussion.

\begin{figure}
\begin{center}
\begin{framed}
\lstset{
  numbers=left,
  numbersep=5pt, 
  basicstyle=\scriptsize\sffamily,
  columns=flexible,
  breaklines=true
  floatplacement={tbp},captionpos=b,
  xleftmargin=8pt,xrightmargin=8pt
}
\begin{lstlisting}[language=haskell]
data (:+:) f g t l = Inl (f t l) | Inr (g t l)
data Term f l = Term (f (Term f)) l
type LangSig = Arith :+: Atom :+: Lit
type LangTerm = Term LangSig
\end{lstlisting}
\end{framed}
\caption{Combining the fragments of  Figure \ref{fig:modularized-example}}
\label{fig:fixed-term}
\end{center}
\end{figure}

\section{Implementation}

\label{sec:implementation}

We have implemented our approach in the \textsc{Cubix} system, named for a fictional robot composed of modular parts that can be reassembled for many purposes. \textsc{Cubix} is organized as a collection of libraries which assist in building incremental parametric syntaxes
and language-parametric transformations. Our implementation totals approximately 13,000 lines of Haskell,
providing support for five languages and several transformations. We build
heavily on the \packagename{compdata} library of
\citet{Bahr2011CompositionalDT} for modular syntax, and
extend it with support for sort injections and the
\packagename{comptrans} library for converting a third-party syntax
definition into modular syntax. We provide generic language components, and modules for labeled terms, control-flow graphs, and higher-order tree traversals.

The code is split between approximately 5000 lines in our language implementations, 2300 in our transformations, 1200 in \packagename{comptrans}, 3900 in our other libraries, and the rest in our driver, miscellaneous code, and minor extensions to 3rd-party libraries. Note that our language implementations do contain a lot of code clones, due to the limits of metaprogramming in Haskell.

The rest of this section discusses \textsc{Cubix} in more
detail. Section \ref{sec:impl-lang} describes implementing an IPS in
\text{Cubix}. Section \ref{sec:impl-transform} describes how to
implement transformations, and Section \ref{sec:impl-elem-hoist} gives
example code. Section \ref{sec:lang-choice} discusses how to
generalize \textsc{Cubix} beyond Haskell and the 5 target languages.

\subsection{Languages}
\label{sec:impl-lang}

\begin{figure}
\centering
\begin{framed}
\begin{longcode}[numbers=left]
data MultiVarDeclAttrsL; data VarInitL; data MultiVarDeclL;
data OptVarInitL; data VarDeclAttrsL; data VarDeclL;
data AssignOpL; data AssignL; data LhsL; data RhsL
data OptVarInit t l where
  JustVarInit :: t VarInitL -> OptVarInit t OptLocalVarInitL
  NoVarInit   :: OptVarInit t OptVarInitL
data VarDecl t l where
  VarDecl :: t VarDeclAttrsL -> t VarDeclBinderL
          -> t OptVarInitL -> VarDecl t VarDeclL
data MultiVarDecl t l where
  MultiVarDecl :: t MultiVarDeclAttrsL ->$\;$ t [VarDeclL]
               -> MultiVarDecl t MultiVarDeclL
data Assign t l where
  Assign :: t LhsL -> t AssignOpL -> t RhsL -> Assign t AssignL
\end{longcode}
\end{framed}
\caption{Generic nodes to model vardecls and assignments}
\label{fig:var-decl-model}
\end{figure}

\begin{figure}
\centering
\begin{framed}
\begin{longcode}[numbers=left]
do let cSortInjections = [QUOTEQUOTECExprIsRhs, QUOTEQUOTEAssignIsCExpr, ...]
   let names = (cSigNames \\ [mkName "Ident", ...])
        ++ cSortInjections ++ [QUOTEQUOTEVarDecl, QUOTEQUOTEP.Ident, QUOTEQUOTEAssign, ...]
   runCompTrans (makeSumType "MCSig" names)
\end{longcode}
\end{framed}
\caption{Generating the incremental parametric syntax}
\label{fig:ips-gen-code}
\end{figure}

As shown in Figure \ref{fig:architecture-diagram}, to add support for a language, the users selects a 3rd-party frontend, and then constructs two derived representations.

\paragraph{Creating the modularized syntax}

There are three steps to creating the modularized syntax. Because the
modularized syntax is identical to the original, but in a different
form, this is completely automatic.

First, for
each algebraic data type in the original AST, the user must create a
language fragment
signature similar to the one in Figure
\ref{fig:modularized-example}. \packagename{comptrans} generates this
code automatically using Haskell's compile-time code-generation engine, Template Haskell
\cite{sheard2002template}. For instance, the command to do this for C
is \code{runCompTrans (deriveMultiComp QUOTEQUOTECTranslationUnit)},
as \code{CTranslationUnit} is the root of the C type.

Second, the user sums these language fragments into a signature for
the language. For C, the command is \code{runCompTrans (makeSumType
  "CSig"$\;$ cSigNames)}. The user may now manually declare types in
terms of \code{CSig}, such as the type of C terms \code{type CTerm =
  Term CSig}, and the signature of labeled C terms \code{type CSigLab
  = CSig :&: Label}.

Finally, another command is used to generate the translations between the \code{language-c} representation and the modularized representation.

\paragraph{Designing a Library of Generic Components} 

Designing a generic language component takes serious thought: it must
be possible to instantiate it in a way that models the corresponding
construct in every language under consideration.

These come in the form of (completely-standalone) manually-written signatures. Figure \ref{fig:var-decl-model} shows \textsc{Cubix}'s definitions for
generic variable declarations and assignments, which we designed to
model the corresponding constructs in C, Java, JavaScript, Lua, and
Python. The empty data declarations like \code{data LhsL} denote
sorts, while the others are generic nodes. This component comes with
many knobs. By providing C-specific nodes of sort \code{VarDeclAttrsL}
and \code{MultiVarDeclAttrsL}, it can model declarations like
\code{const int x = 1, *y;}. By providing empty nodes of those sorts,
it can model Lua and Python variable declarations.

\paragraph{Creating the IPS}


The user must now decide how to instantiate the generic components in
Figure \ref{fig:var-decl-model} to model their language-specific
counterparts. For instance, in C, assignments are expressions, and
expressions are assignment right-hand sides. The user specifies this
by generating a sort injection from \code{AssignL} to \code{CExprL}
and from \code{CExprL} to \code{RhsL}, and does similar for
\code{LhsL} and \code{AssignOpL}.

The user is now ready to define the IPS for the language. This is done
by expressing the old signature as a compile-time list of symbols, and
literallly removing the language-specific components and adding the
generic ones. Figure \ref{fig:ips-gen-code} gives the code used to generate the IPS C signature, \code{MCSig}

Finally, the user must write a translation from the modularized syntax
to the IPS. They need only write code for the cases where the syntaxes
differ, i.e.: to replace language-specific nodes with generic ones.

This completes the process depicted in Figure \ref{fig:architecture-diagram}.

\paragraph{Other support}

Some transformations may require other language-infrastructure, such
as a control-flow graph generator. The IPS representation makes it easy to share code across languages; our 5 CFG-generators average 101 LOC.

In our experience, creating an incremental parametric syntax for a new language takes 1-2 days. We have implemented support for C, Java,
JavaScript, Lua, and Python, using the parsers, pretty-printers,
and syntax definitions from the Haskell libraries \packagename{language-c} \cite{authors2016language-c-0.5.1},  
\packagename{language-java} \cite{broberg2015language-java-0.2.8}, \packagename{language-javascript} \cite{zimmerman2016language-javascript-0.6.0.9}, \packagename{language-lua} \cite{ağacan2016language-lua-0.10.0}, and lastly \packagename{language-python} \cite{pope2016language-python-0.5.4}. Because of problems with the parser for \packagename{language-java}, we instead use a Java parser written in Java, the \url{javaparser.org} parser \cite{javaparser}, and translate its results into the \packagename{language-java} AST. Despite their names, these libraries were all implemented independently by different authors, and share no common infrastructure beyond standard libraries. We fixed bugs in all of their pretty printers but were otherwise not involved with their development. Some of our fixes have yet to be merged upstream.


\subsection{Transformation Support}
\label{sec:impl-transform}

\setlength{\belowcaptionskip}{4pt}
\begin{figure}
\centering
\begin{framed}
\begin{longcode}[numbers=left]
type HasSyntax f = (VarDecl :<: $\;$ f, MultiVarDecl :<: $\;$ f
  , OptVarInit :<: $\;$ f, Ident :<: $\;$ f, Assign :<:$\;$ f, AssignOpEquals :<: $\;$ f
  , Block :<: f, ListF :<: $\;$ f, ExtractF [] (Term f))
type CanHoist f = (HasSyntax f, VarInitToRhs (Term f)
  , VarDeclBinderToLhs (Term f), HTraversable f
  , InjF f MultiVarDeclL BlockItemL, InjF f AssignL BlockItemL)
\end{longcode}
\end{framed}
\caption{Constraints for the elementary hoist transform}
\label{fig:hoist-constraints}
\end{figure}
\resetcaptionlengths

A language-parametric transformation makes limited assumptions about its target language. This is done by parameterizing the transformation over operations on the nodes and terms of the language, given in the form of Haskell typeclasses.

The constraints for the elementary hoisting transformation, \code{CanHoist} in Figure \ref{fig:hoist-constraints}, depicts the full spectrum of such operations. The elementary hoisting transformation can run on any language that satisfies these constraints, and gives a compile error on any that do not. First, there are constraints that the language contain generic nodes. This is given as the $\prec$ constraint from \packagename{compdata}, which provides an injective function from the generic node to terms of the language, \code{inject}, and its partial inverse, \code{project}. As a second class, the \code{InjF} constraints are sort injections as discussed in Section \ref{sec:sort-injections}. Finally, \code{VarDeclBinderToLhs} and \code{VarInitToRhs} provide the language-specific operations of elementary hoisting, discussed in Section \ref{sec:building-hoisting}. Overall, these constraints allow a transformation to make a limited set of assumptions about its target languages, allowing it to handle the intricate details of many languages while maintaining a high level of generality.

There are also a couple more technical constraints. The interface \code{HTraversable} from \code{compdata} interface offers generic tree-traversal operations. \code{MaybeF} and \code{ListF} provide tree nodes representing optional nodes and lists of nodes, so a node representing a list of block items may have sort \code{[BlockItemL]}. There are then operations \code{extractF} and \code{insertF} to convert between values of type \code{Term f [l]} (term of sort ``list of \code{l}'') and values of type \code{[Term f l]} (list of terms of sort \code{l}).

We have built a library of \emph{strategy combinators} \cite{lammel2002typed} called \packagename{compstrat}. With strategy combinators, the user can turn a set of single-node rewrites into a complicated traversal pattern in a single line of code. \packagename{compstrat} provides similar functionality to other strategy combinator libraries such as Scrap Your Boilerplate \cite{lammel2003scrap}, Strafunski \cite{lammel2003strafunski}, and KURE \cite{gill2009haskell}.

We have also built miscellaneous other infrastructure to support our
transformations. The most interesting of these is the control-flow
based inserter. Inserting a statement before a loop condition causes it to be placed before the loop, before the end of the loop,
and before every \code{continue} statement.

\subsection{Example: Implementing the Elementary Hoisting Transformation}
\label{sec:impl-elem-hoist}

\begin{figure*}
\centering
\begin{framed}
\begin{longcode}[numbers=left]
declToAssign :: (CanHoist f) => Term f MultiVarDeclAttrsL -> Term f VarDeclL -> [Term f BlockItemL]
declToAssign mattrs (VarDecl' lattrs b optInit) = case optInit of
  NoVarInit'        -> []
  JustVarInit' init -> [injF (Assign' (varDeclBinderToLhs b) AssignOpEquals' (varInitToRhs mattrs b lattrs init))]

removeInit :: (CanHoist f) => Term f VarDeclL -> Term f VarDeclL
removeInit (VarDecl' a n _) = VarDecl' a n NoVarInit'

splitDecl :: (CanHoist f) => Term f BlockItemL -> ([Term f BlockItemL], [Term f BlockItemL])
splitDecl (projF -> (Just (MultiVarDecl' attrs decls)))
          = ([injF (MultiVarDecl' attrs (mapF removeInit decls))], concat (map (declToAssign attrs) (extractF decls)))
splitDecl t = ([], [t])

hoistBlockItems :: (CanHoist f) => [Term f BlockItemL] -> [Term f BlockItemL]
hoistBlockItems bs = concat decls ++ concat stmts
  where (decls, stmts) = unzip (map splitDecl bs)

elementaryHoist :: (CanHoist f) => Term f l -> Term f l
elementaryHoist t = transform hoistInner t
  where hoistInner :: (CanHoist f) => Term f l -> Term f l
        $\;$hoistInner (project -> (Just (Block bs e))) = Block' (liftF hoistBlockItems bs) e
        $\;$hoistInner t                                = t

\end{longcode}
\end{framed}
\caption{Implementation of the elementary hoist transformation}
\label{fig:elem-hoist-code}
\end{figure*}

This section presents the full implementation of the elementary hoisting transformation from Section \ref{sec:building-hoisting}. Figure \ref{fig:elem-hoist-code} gives the code; Figure \ref{fig:hoist-constraints} showed the \code{CanHoist} constraint. We omit the 30 lines of code giving the three language-specific instances of \code{VarInitToRhs} and \code{VarDeclBinderToLhs}.

The code implements the algorithm described in Section
\ref{sec:building-hoisting}. Execution begins at
\code{elementaryHoist}, which runs \code{hoistBlockItems}
over every block. It does so by using \packagename{compdata}'s \code{transform} function to run \code{hoistInner} over every node, which uses \code{project} to test if a node is a block. Later, the sort injections are used via \code{projF} and \code{injF} to operate on the subset of \code{BlockItem}'s that the transformation knows about.

In this example, we have tried to avoid many of the vagaries of
Haskell syntax as well as more advanced features of \textsc{Cubix}. Nonetheless, some advanced features are
present. The sum-of-signatures approach distinguishes between nodes, which may lie in an arbitrary AST, and terms, which are tied to a single language. The vanilla data constructors
of Figure \ref{fig:var-decl-model} like \code{Assign} construct nodes of a
signature fragment, while their primed variants like \code{Assign'} construct and pattern match on terms. We explained the
\code{extractF} and \code{insertF} functions in Section
\ref{sec:impl-transform}; these are used to implement the \code{liftF}
and \code{mapF} functions, which are used to operate on trees of type
\code{Term f [l]}. Finally, the syntax \code{f (view -> Just x) = ...} is a Haskell
\emph{view pattern} \cite{wadler1987views} which is syntactic sugar
for \code{f t = case view t of Just x -> ...}, with pattern match
failure proceeding to the next case. 

\subsection{Choices of Target and Implementation Languages}

\label{sec:lang-choice}

When we speak about \textsc{Cubix}, we always find people who want to use it or something like it for their applications. What would it take to implement a system like \textsc{Cubix} in a different language? And what about supporting other languages, such as ML or Prolog or Haskell itself?

\paragraph{What do we gain from these fancy types?}

\begin{table}
\begin{center}
\caption{Various term types in \textsc{Cubix}}
\label{table:cubix-types}
\begin{tabular}{|l|l|}
\hline
\textbf{Type signature} & \textbf{Description} \\
\hline
\code{Term f AssignL} & Assignments in any language \\
\hline
\code{Term MJavaSig l} & Java terms of any sort \\
\hline
\code{(Assign :<: f) => Term f IdentL} & An identifier in any language
                                        that contains generic
                                        assignments \\
\hline
\code{Term f (StatSort f)} & \makecell[cl]{A statement in any
                             language. The statement sort is \\  language-specific.} \\
\hline
\makecell[cl]{
\code{(InjF f IdentL PositionalArgExpL} \\ \code{, CallAnalysis f) => Term f
  IdentL}} & \makecell[cl]{An identifier in any language which supports a call
            analysis, \\ and where identifiers may be used as ordinary
            arguments \\ to functions} \\
\hline
\end{tabular}
\end{center}
\end{table}

\textsc{Cubix}'s implementation of incremental parametric syntax uses
some rather advanced type system features. Case in point: the current
implementation uses a total of 32 GHC extensions. What's the
benefit of all this, and can it be replicated in a language other than Haskell?

There are two primary benefits. The first is the precise typing. The second is dispatch: the compiler
uses the type information to choose appropriate language-specific and
sort-specific operations. Both are indispensible for building
multi-language tools. And their synthesis allows generic programming.

Consider the example types in Table \ref{table:cubix-types}. They show
how, using the \code{Term f l} type, developers can restrict
operations to certain sorts of terms, languages, and properties of the
language. These restrictions are all enforced by the compiler.

Without these types, it's still quite easy to write a function that
can accept many kinds of terms, by giving them all a single ``Node''
type. This is the dynamically-typed or ``generic node'' approach, used
in many language workbenches. This is not enough to get
language-parametric transformation, as removing the types does not
remove the network of constraints between AST nodes. There must be the
extra step of converting part of the tree to some common form, as done
in IPS.  And these conversions introduce massive room for errors.

Our own experience attempting this kind of generic programing in
JavaScript, as well as fixing type errors during normal \textsc{Cubix}
development, makes us pessimistic about trying it without
precise types. It's far too easy to e.g.: attempt to use an assignment as an
expression, when that is not legal in every language.

The second major benefit is dispatch. Consider writing a transformation that works on any language
where functions can be turned into lambdas. If we were to implement
this transformation in a language without typeclasses, we would make the
transformation take a ``turn function into lambda'' operation as a
parameter. This operation would then need to be transitively supplied
to every piece of the transformation that needed it.

Conversely, in Haskell, we'd simply add a condition to the constraints
for the transformation, as in Figure \ref{fig:hoist-constraints}, and
it would be propagated to all components. And when attempting to call this transformation
on a specific language, the compiler will automatically find and
supply the correct instance of the ``turn function into lambda'' operation.

In other words, it is very easy to write a generic operation that
includes language-specific pieces. Doing this at a smaller level is generic
programming. For instance, \code{subterms e :: [Term f IdentL]} gives
all identifiers contained in \code{e}. The equivalent code in a
language without typeclasses would be something like \codebasic{subterms(e, Filters.checkSort(SORT_IDENT))}. Meanwhile, Haskell automatically
supplies \code{subterms} with the correct sort-classification check by
looking at the types. So, this representation is useful even when only working with one
language.

\paragraph{IPS in other languages} 

Our implementation of incremental parametric syntax relies heavily on
three distinctive features. We discussed the use of type classes
above. The \code{Term f l} type relies on GADTs to work (else
\code{Term f IdentL} could not be fundamentally different from
\code{Term f AssignL}). And we've used Haskell's built-in code
generation, Template Haskell, throughout this paper. Haskell is the
only language we know of that supports all three features. But even
Haskell is not a perfect language for building this kind of system,
and we still have a wishlist of language features that would make
\textsc{Cubix} development much easier (e.g.: pattern matching that
works better with modular syntax).

We can envision a \textsc{Cubix}-like framework in a language
without any of these three features. It would use an extra-linguistic
tool to generate boilerplate code. Users would pass in operations
manually in lieu of typeclasses, at some inconvenience. But without
GADTs, we see only two options, neither of them appealing: either
write a custom type-checker/analyzer, or face the pitfalls of
dynamically-typed terms.

\paragraph{Supporting more languages}

To share code between languages, these languages must have nodes which are similar enough to design a generic node whose semantics model all of them. Expressions are similar in C and ML, and we see no barrier writing transformations that can operate on them both. On the other hand, the execution semantics of Prolog nodes differs substantially from imperative and functional languages, and so we do not expect to be able to write C/Prolog transformations.

Integrating \textsc{Cubix} with a third-party parser only requires that the parser output to a Haskell ADT. We hence picked languages that already had good Haskell libraries, but it can integrate with parsers written in other languages by writing a wrapper, as we have done for Java.

\section{Evaluation}

In the previous section, we argued that the insights of \textsc{Cubix} make language-parametric transformations easy to write. In this Section, we demonstrate a realistic language-parametric transformation and its application to real software, and further evaluate the following two claims:

\begin{itemize}
\item \emph{Readability}: These transformations produce readable output, similar to what a human would write. They do not needlessly destroy the program's structure, as do IR-based transformations.
\item \emph{Correctness}: Despite the low effort needed per language, transformations can maintain correctness even when faced with the intricacies of multiple languages.
\end{itemize}

Additionally, because we built the tool of Section \ref{sec:ipt} after the rest of the work in this paper, our experience building also supports our claim that it is easy to extend an IPS as more features are needed to support new transformations.

\subsection{A Realistic Whole-Program Refactoring}

\label{sec:ipt}

In this section, we present the IPT tool ({\bf i}nterprocedural {\bf p}lumbing {\bf t}ransformation) for threading variables through chains of function calls, inspired by the Dropbox and Facebook stories in Section \ref{sec:has-dropbox-story}. We built the IPT tool as a language-parametric transformation which we developed simultaneously for all 5 languages supported by \textsc{Cubix}.

The IPT tool takes a method and a parameter name, and recursively has
all callers pass down said parameter, asking for user approval for
each change.  Figure \ref{fig:ipt-example} presents a scenario where
the end goal is to replace the call to \code{strcpy} within \code{f1}
with \code{strncpy}, and the barrier is that the programmer is missing
the \code{len} parameter needed by \code{strncpy}. He invokes the IPT
tool to add an \code{int len} parameter to \code{f1}, pressing ``Yes''
4 times to change lines 1, 5, 4, and 9, so that the existing
\code{len} parameter in \code{f3} is passed through 2 layers of
function calls to \code{f1}. After the IPT tool is done, the
programmer can now manually change line 2 to \code{strncpy(buf,
  "Hello", len);}. The IPT tool has automated plumbing data through
the system; all that is left for the programmer is to choose where it
comes from, and how it's used.

\begin{figure}
\begin{center}
\lstset{
  numbers=left,
  numbersep=5pt, 
  basicstyle=\scriptsize\sffamily,
  columns=flexible,
  breaklines=false
  floatplacement={tbp},captionpos=b,
  xleftmargin=8pt,xrightmargin=8pt
}
\begin{tabular}[t]{|p{0.3\textwidth}|p{0.3\textwidth}|}
\hline
\begin{lstlisting}[language=c]
void f1(char* buf) {
  strcpy(buf, "Hello");
}
void f2(char* buf) {
  f1(buf);
}
void f3(int len) {
  char *buf = malloc(len);
  f2(buf);
}
\end{lstlisting}
 &
\begin{lstlisting}[language=c]
void f1(char* buf, int len) {
  strcpy(buf, "Hello");
}
void f2(char* buf, int len) {
  f1(buf, len);
}
void f3(int len) {
  char *buf = malloc(len);
  f2(buf, len);
}
\end{lstlisting} \\
\hline
\end{tabular}
\caption{Input/output example of the IPT tool on C}
\label{fig:ipt-example}
\end{center}
\end{figure}

Building this tool also shows that it is easy to extend an IPS to
support new transformations. We had not needed a generic notion of
functions for our previous transformations we implemented (see Section
\ref{sec:benchmark-transformations}), so we made an incremental change
to our parametric syntax. In 3 hours, we designed a generic fragment
for function definitions and calls that could be instantiated to model
the features of all languages under consideration. It took us 21 hours
to design and implement the changes to all 5 language representations,
proportional to the complexity of each language  (e.g.: 5 hours to
understand C declarations and their many variations). We thus obtained
the incrementality benefits of IPS: we didn't need to build up-front support for functions, but could still build transformations that needed them, and we now have support for functions for all future transformations.

With the generic syntax for functions in place, building the tool itself took only 19 hours. Altogether, the extensions to \textsc{Cubix} averaged 5 hours per language, while the tool itself averaged another 4 hours per language.

The implementation is fairly straightforward: it maintains two queues of function calls and definitions to be modified, and prompts the user about each potential change. After modifying each function definition, it uses a static analysis to find all callers and add them to the queue. This static analysis is a parameter of each language, but the analyses for each language may use a shared implementation using techniques of multi-language analysis. Making this analysis more precise means the user will be prompted for fewer erroneous changes.

Our IPT tool is still a prototype, with a minimal command-line UI, an imprecise call analysis, and incomplete support for C function prototypes. Nonetheless, we have used it in three real case studies in Java and Python, in addition to toy programs in the other three languages.

We first used it on SimpleDB \cite{simpledb}, a teaching database used at several universities. SimpleDB totals 23,000 lines of Java, and 11,500 lines of tests. It frequently accesses a global \code{BufferPool} object by calling \codebasic{Database.getBufferPool()}. We used a two-step process to eliminate this global. First, we used the IPT tool to thread a \code{bufferPool} parameter throughout the program. This changed all \codebasic{Database.getBufferPool()} calls to instead read \codebasic{Database.getBufferPool(bufferPool)}. Second, we applied a find-and-replace to the entire program to simplify them to \code{bufferPool}. We then manually changed entry points to the program to supply this \code{bufferPool} parameter. Altogether, the IPT tool modified 484 lines across 41 files, while we manually modified 50 lines across 24 files. All tests pass.

We then did two smaller case studies in Python. Flask \cite{grinberg2014flask} is a Python web micro-framework which totals 6500 lines of Python and 5700 lines of tests. We used the IPT tool to modify the \codebasic{_get_exc$\_$class$\_$and$\_$code} function to take a default exception code, and propagated this parameter up several layers. The IPT tool modified 21 lines across 2 files. We only manually changed 2 lines: to use this parameter, and to supply it at the top of the chain. Tornado\cite{dory2012introduction} is a Python web server owned by Facebook. It comprises 22,000 lines of Python and 16,000 lines of tests. We changed the \code{is$\_$valid_ip} function to take an \code{accept_ipv6} parameter, and propagated this parameter up several layers. The IPT tool changed 22 lines across 9 files. We used a find-and-replace to provide a default value to many new parameters, and then changed 3 lines manually. All tests pass for both Flask and Tornado.

For all five languages, we also tested the IPT tool on a toy program consisting of three functions across three files that call each other; the tool successfully propagated a parameter through all three functions.


\subsection{Benchmark Transformations}
\label{sec:benchmark-transformations}

To more rigorously evaluate our system, we have implemented three smaller source-to-source transformations. These were chosen to explore the space of operations used by program transformations and to require a minimum of user input. Table \ref{table:transformations} lists them and their line counts.

\begin{itemize}
\item The hoisting transformation \textbf{Hoist}, which lifts variable declarations to the top of their scope. This is similar to elementary hoisting in Sections \ref{sec:building-hoisting} and \ref{sec:impl-elem-hoist}, except that it also supports Lua, and uses additional machinery to avoid hoisting shadowed variables, and to deal with language special-cases such as C's structure initializers. Figure \ref{fig:hoisting-example} gave a mundane example; Figure \ref{fig:hoisting-js} gives an example handling a JavaScript special case. This transformation supports all languages except Python, which lacks variable declarations.

\item The test-coverage instrumentation transformation
  \textbf{Testcov}, which prefixes each basic block in the source code
  with an assignment which marks that that block has executed. This
  produces data that could be fed into a test coverage tool. It was inspired a Semantic Designs
  tool which implements this transform separately for a dozen
  languages \cite{sd-testcoverage}. This transformation supports all
  languages. Figure \ref{fig:testcov-java} shows an example special case for Java.
\item The three-address code transformation \textbf{TAC} hoists all nested computations into temporary variables, e.g.: changing \code{1+1+1} into \code{t=1+1; t+1}. This is a deceptively complicated transformation, difficult to write at the source level for even one language. Figures \ref{fig:tac-js} and \ref{fig:tac-python} show a few of the complexities it supports, all handled cleanly by \textsc{Cubix}'s general infrastructure for operator strictness and CFG-based insertion. This transformation  supports JS, Lua, and Python. It does not support Java or C because declaring the temporary variables would require type inference, which in turn requires symbol-table construction, a heavyweight piece of language infrastructure.
\end{itemize}

\begin{figure}
\begin{center}
\lstset{
  numbers=left,
  numbersep=5pt, 
  basicstyle=\scriptsize\sffamily,
  columns=flexible,
  breaklines=false
  floatplacement={tbp},captionpos=b,
  xleftmargin=8pt,xrightmargin=8pt
}
\begin{tabular}[t]{|p{0.2\textwidth}|p{0.20\textwidth}|}
\hline
\begin{lstlisting}[language=javascript]
function f() {
  "use strict";
  if (x) {
      var y =1;
  }
}
\end{lstlisting}
 &
\begin{lstlisting}[language=javascript]
function f() {
  "use strict";
  var y;
  if (x) {
    y = 1;
  }
}
\end{lstlisting} \\
\hline
\end{tabular}
\caption{Hoisting JavaScript, showing interactions with JS's \code{"use strict";} pragmas and lack of inner scopes. No JS-specific hoisting code is needed, only a precise representation of JS blocks.}

\label{fig:hoisting-js}
\end{center}
\end{figure}

\begin{figure}
\begin{center}
\lstset{
  numbers=left,
  numbersep=5pt, 
  basicstyle=\scriptsize\sffamily,
  columns=flexible,
  breaklines=false
  floatplacement={tbp},captionpos=b,
  xleftmargin=8pt,xrightmargin=8pt
}
\begin{tabular}[t]{|p{0.4\textwidth}|p{0.4\textwidth}|}
\hline
\begin{lstlisting}[language=java]
public static void foo(int x) {
  if (x > 0) {
    while(true)
      x++;
    // Unreachable code
  }
}
\end{lstlisting}
 &
\begin{lstlisting}[language=java]
public static void foo(int x) {
  TestCoverage.coverage[0] = true;
  if (x > 0) {
      TestCoverage.coverage[1] = true;
      while (true)
        x++;
  }
  TestCoverage.coverage[2] = true;
}

\end{lstlisting} \\
\hline
\end{tabular}
\caption{Test coverage for Java. A naive transformation would insert a test coverage statement on line 5 after the while loop, causing an "unreachable code" compile error. This case is supported purely through the CFG-generator, requiring no Java-specific code in the transformation itself.}
\label{fig:testcov-java}
\end{center}
\end{figure}

\begin{figure}
\begin{center}
\lstset{
  numbers=left,
  numbersep=5pt, 
  basicstyle=\scriptsize\sffamily,
  columns=flexible,
  breaklines=false
  floatplacement={tbp},captionpos=b,
  xleftmargin=8pt,xrightmargin=8pt
}
\begin{tabular}[t]{|p{0.2\textwidth}|p{0.20\textwidth}|}
\hline
\begin{lstlisting}[language=javascript]
while (f() && g(1+1)) {
    x++;
}

\end{lstlisting}
 &
\begin{lstlisting}[language=javascript]
var t1 =  f();
var t2;
if (t1) {
  var t3 =  1 + 1;
  t2 = g(t3);
}
var t4 =  t1 && t2;
while (t4) {
  x++;
  t1 = f();
  if (t1) {
    var t3 =  1 + 1;
    t2 = g(t3);
  }
  t4 = t1 && t2;
}
\end{lstlisting} \\
\hline
\end{tabular}
\caption{TAC transformation example for JavaScript, showing handling of loops and non-strict operators.}
\label{fig:tac-js}
\end{center}
\end{figure}

\begin{figure}
\begin{center}
\lstset{
  numbers=left,
  numbersep=5pt, 
  basicstyle=\scriptsize\sffamily,
  columns=flexible,
  breaklines=false
  floatplacement={tbp},captionpos=b,
  xleftmargin=8pt,xrightmargin=8pt
}
\begin{tabular}[t]{|p{0.2\textwidth}|p{0.20\textwidth}|}
\hline
\begin{lstlisting}[language=python]
if x is None:
  doThing1()
elif x.foo():
  doThing2()
\end{lstlisting}
 &
\begin{lstlisting}[language=python]
t1 = x is None
if t1:
    del t1
    doThing1()
else:
    del t1
    t2 = x.foo()
    if t2:
        del t2
        doThing2()
    else:
        del t2
\end{lstlisting} \\
\hline
\end{tabular}
\caption{TAC transformation example for Python. It avoids computing \codebasic{x.foo()} when \code{x} is \code{None}, and deletes all temporaries immediately after use, as Python is sensitive to the GC behavior. Adding the \code{del} statements is 5 lines of Python-specific code.}
\label{fig:tac-python}
\end{center}
\end{figure}

\begin{table*}
\begin{center}
\caption{Transformations implemented and their size. Line counts are
  split into the core code of the transformation, plus the
  per-language code to support language-specific operations and
  customization. Line counts exclude the file prologue, i.e.: they
  count from the first line of code which is not an \code{import}
  statement.}
\label{table:transformations}
\begin{tabular}{l|lrr}
\hline
{\bf Transformation} & {\bf Languages Supported} & {\bf Core LOC} & {\bf Extra LOC per language} \\
\hline
Hoist & C, Java, JavaScript, Lua & 154 & 65 \\
Testcov & All & 77 & 25 \\
TAC & JavaScript, Lua, Python & 360 & 116 \\
\hline
\end{tabular}
\end{center}
\end{table*}

\noindent 

All transformations use a mixture of generic and language-specific code. However, the language-specific code is usually much less complex, and fewer lines are needed per-language, as shown in Table \ref{table:transformations}.

\subsection{Readability Study}

Transforming through an IR mutilates the program, but transforming
with incremental parametric syntax preserves information. To prove
this, we ran a ``Turing test,'' where our system and human programmers
transformed code in the same manner, and judges from Mechanical Turk
rated them both on readability. This section presents highlights of
this study; full details are given in Appendix \ref{app:study-long}.
Our study aimed to prove the following hypothesis for each language:

\begin{hyp}
\label{hyp:noninf}
\item For a random code sample and our transformations, a human judge will rate the machine-transformed code at most 1  worse on a 1-5 scale than the human-transformed code in expectation.
\end{hyp}

Note that this study was completed using earlier versions of the transformations which failed some tests.

\subsubsection{Phase 1: The RWUS Suite}

\label{sec:describes-rwus}

As objects in our study, we needed (1) representative samples of real-world code, and (2) an objective measure of whether a transformed sample was equivalent to the original. As random samples of code do not come with thorough tests, we created our own.

The RWUS (Real World, Unchanged Semantics) suite consists of 50 functions across 5 languages randomly selected from top GitHub projects, together with a test suite designed to catch any semantic changes to the program. Each function is distributed as a file that can be compiled and executed without any dependencies. Our test cases have full path coverage and ensure all mocked functions are called in the expected order with the expected arguments. The tests are incredibly thorough: while the actual samples total 1158 lines of code, the RWUS suite totals 8070 lines of code.

We expect the RWUS suite to be useful in testing other semantics-preserving transformations. It is available from:

\begin{center}
\url{https://github.com/jkoppel/rwus}
\end{center}

\subsubsection{Phase 2: Human-Written Transformations}

In the next phase, we recruited programmers, gave them random entries from the RWUS suite, and had them perform each of our transformations by hand on the function. After normalizing formatting, 24 of the 120 resulting programs were identical to the machine-transformed  ones. Another 7 of the machine-transformed ones failed their tests; the remaining programs were sent to human judges for Phase 3.


\subsubsection{Phase 3: Mechanical Turk}

In Phase 3, we asked human judges from Mechanical Turk to rate the manually-transformed code from Phase 2 along
with their automatically transformed counterparts. In random order, they were given an original program and its human- and machine-transformed versions, and asked to rate both on a 1-5 scale, prioritizing correctness, then whether the transformation was done correctly, and third on code quality. We also employed measures to catch unqualified and inattentive Turkers and discard their data, chiefly canary questions (e.g.: a pair of identical programs which should be rated identically). We assigned each of the 20-30 transformed samples to 10 judges, giving us up to 300 ratings per language.


\subsubsection{Results}

\begin{figure*}
\begin{center}
\includegraphics[scale=0.4]{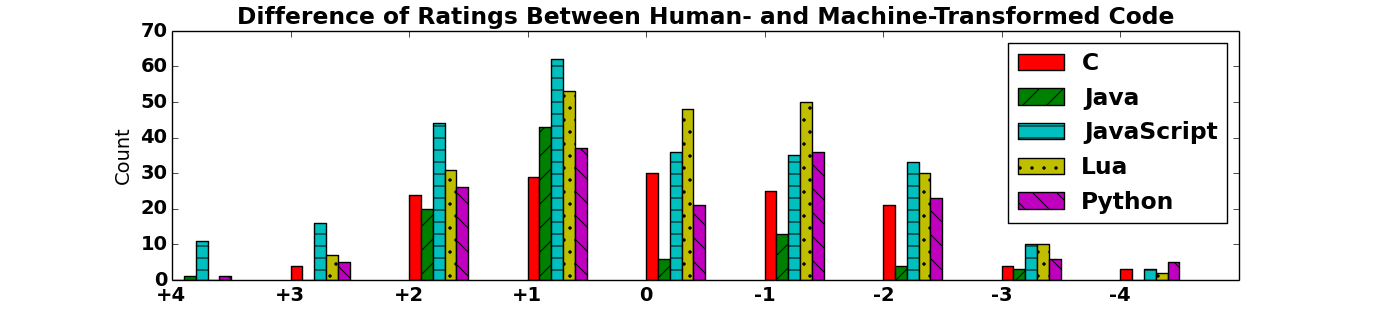}
\end{center}
\caption{Counts of differences between the ratings of the machine transformations and the human transformations. The leftmost bars represent cases where the judge rated the machine-produced output higher than the human-produced.}
\label{fig:human-study-results}
\end{figure*}

For each language, we tabulated the difference in ratings between the human-written and automatically transformed programs. Our results are given in Figure \ref{fig:human-study-results}. The average differences in ratings ranged from $-0.075$ for Python (favoring the humans) to $+0.633$ for Java (favoring the machine). The differences for C, JavaScript, and Lua were $-0.014$, $+0.396$, and $-0.052$ respectively.

We test Hypothesis \ref{hyp:noninf} using the techniques of non-inferiority testing \cite{wellek2010testing}. We combined the data from the human judges with the samples that did not get sent to Phase 3. For the samples where the human- and machine-transformed versions, were identical, we included them as if the humans had rated them identically; for the ones that failed their tests, as if rated to maximally penalize the machine. We then tested each of the $5$ hypotheses using a paired t-test. For each language, it showed that the machine-transformed code was non-inferior by a non-inferiority margin of at most $1$ with $p<10^{-8}$. In retrospect, this data had the power to prove the hypothesis with a much smaller non-inferiority margin.

Considering both the raw data and the statistical tests, our study provides strong evidence that the output of transformations in \textsc{Cubix} is no less readable than hand-transformed code, showing that implementing source-to-source transformations with incremental parametric syntax avoids the mangling common to IR-based approaches.

\subsection{Correctness}

\label{sec:evaluation-correctness}

We claim it's feasible to write semantics-preserving language-parametric transformations with our approach. Hence, we collected language test suites for each of the 5 languages, and improved our transformations until we had a $100\%$ pass rate for all transformations on all languages.

The caveat, though, is that there are some tests which the transformations should not pass. First, we use third-party parsers and pretty-printers, all of which have bugs. We contributed some bug fixes to all of these projects, but issues still remain. Second, all of the dynamic languages have self-referential tests which will never pass (e.g.: "assert this function was declared on line 37"). We rule out these cases by first checking if the test still passes after running the identity transformation \textbf{Ident}, consisting of parsing and pretty-printing the program. $93.4\%$ of tests pass this \textbf{Ident} transformation. This discussion excludes the Lua test suite, which has other issues explained below.

Table \ref{table:versions} lists the language implementations and test suites used in our evaluation. The C, Lua, and Python tests come from their implementations, while the JavaScript ones come from the official specification conformance test suite. The authors of K-Java, \citet{bogdanas2015k}, report that no Java language tests are publicly available, and hence created their own specification tests, which we use. We restricted ourselves to the core language tests of \packagename{test262}, using the same subset as the JavaScript semantics KJS \citep{park2015kjs}, and omitted a small handful of multi-file Java tests among the Java ones, which caused problems with our test harness. We used the entirety of the Lua, Python, and C test suites.

Table \ref{table:test-results} shows the number of passing tests for
each language and transformation. The \textbf{Ident} transformation is
a baseline transformation which simply parses and pretty prints a
program, in order to filter out ``bad tests'' as described above. The
\textbf{Hoist}, \textbf{Testcov}, and \textbf{TAC} columns show the
results of their respective transformations. Comparing the other
transformations to \textbf{Ident}, all but $12$ tests pass. The
failing JavaScript and Python tests are all self-referential tests
that were not ruled out by \textbf{Ident}. The failing JavaScript
tests all use \code{function.toString()}, which retrieves the textual
source code of the function. The Python ones inspect the runtime representation of functions, such as the presence of opcodes in the compiled bytecode or the number of bytes used per stack frame. The failing Java hoist test is actually due to a crash of $\packagename{javac}$.  Manual inspection shows that this program is indeed correct, and the bug has been confirmed by the JDK developers \cite{jdk-bug}.

\begin{table}
\begin{center}
\caption{Compilers/interpreters and test suites used in evaluation}
\label{table:versions}
\begin{tabular}{l|llrr}
\hline
{\bf Language} & {\bf Compiler/Interpreter} & {\bf Test Suite} & {\bf Test Files} & {\bf Total Test LOC} \\
\hline
C &  GCC 6.3.0\_1 & gcc-torture \nocite{gcc-torture} & 1394 & 53,637 \\
Java & JDK 1.8.0\_65 & K-Java \nocite{bogdanas2015k} & 755 & 26,568 \\
JS & Node.js v0.10.24 & test262 \nocite{test262} & 2782  & 128,698 \\
Lua & Lua.org 5.3.3 & Lua Tests \nocite{lua-tests} & 28 & 12,017 \\
Python & CPython 3.7.0a0 & CPython Tests \nocite{python-tests} & 404 & 249,499  \\ 
\hline
\end{tabular}
\vspace{3pt}
\end{center}
\end{table}

While we were very successful with the other language test suites, we found substantial barriers using the Lua test suite to test our transformations. Its tests are highly self-referential, including a check that the ``test'' function is defined on line 17, points where it undefines every global variable, and tests that break if a file changes character encoding. We nonetheless tried.

As the Lua tests are distributed as a single program, we modified the Lua test suite to maintain a count of passed assertions, instead of stopping at the first failure, and deleted
some of the overly self-referential assertions. We found that the total number of calls to
\code{assert} was nondeterministic, but the number of failing
assertions was not. In one set of runs, we obtained the following
numbers: $70440 / 70456$ passing assertions for the original, $70279 /
70295$ for the identity transformation, and $70463 / 70479$ for
hoisting. We gave up attempting to get it working for the test
coverage transform, due to crashes related to its metaprogramming
around global variables. We similarly gave up for the TAC transformation, because the Lua VM does not allow for more than 200
local variables in any scope, and the TAC transformation overwhelms this easily. We conclude that the Lua test suite is unsuitable for testing program transformations.

\begin{table}
\begin{center}
\caption{Results of each transformation on the test suites}
\label{table:test-results}
\begin{tabular}{l|rrrrr}
\hline
{\bf Lang} & {\bf Total } & {\bf Ident} & {\bf Hoist} &  {\bf Testcov}
  & {\bf TAC}\\
\hline
C & 1394 & 1305 & 1305 & 1305 & N/A \\
Java & 755 & 745 & $^*744$ & 745 & N/A \\
JS & 2782 & 2573 & 2573 & 2568 & 2572  \\
Python & 404 & 360 & N/A & 358 & 357 \\
\hline
Lua & \multicolumn{4}{c}{Reported separately} \\
\hline
\multicolumn{5}{l}{* Not including test which crashed \packagename{javac}} \\
\hline
\end{tabular}
\end{center}
\vspace{2pt}
\end{table}

\section{Related Work}

Our work is most directly based on the data types \`{a} la carte
approach to modular syntax \cite{swierstra2008data}, and its
extensions in work on compositional data types by
\citet{Bahr2011CompositionalDT}. The extension to multi-sorted terms
was introduced in \citet{yakushev2009generic}. Other approaches to
modular syntax include tagless-final \cite{kiselyov2012typed}, object
algebras \cite{Zhang2015ScrapYB}, and modular reifiable matching
\cite{oliveira2016modular}. All these works share the same limitation:
supporting a language requires building it from scratch in terms of
special components. We overcame this limitation by using
sort-injections to intermix a generic representation with one from
existing frontends.
 We previously described \textsc{Cubix} in a
poster-paper in OOPSLA 2017 \citep{cubixPoster}.

This work on modular syntax is joined by work on modular semantics,
such as modular monadic semantics \cite{liang1995monad} and its cousin
modular monadic meta-theory \cite{Delaware2013ModularMM}, as well as
modular SOS \cite{Mosses2004ModularSO} and its successor work on
funcons \cite{churchill2015reusable}. These are used to build and
verify interpreters for multiple languages, and will likely be
necessary to extend our work to verifying multi-language transformations.

2003 saw a Dutch grant on language-parametric refactoring
\cite{dutchgrant}, building on a prototype by L\"{a}mmel
\cite{lammel2002towards,heering2004generic}. Their approach was to
parameterize a transformation on (1) a fixed number of
(language-specific) sorts used by the transformation, and (2) a set of
primitive transformation operations, given as functions over these
sorts. In this approach, the ASTs are opaque to the generic code, and hence
essentially all computation happens in the language-specific functions. Conversely, in our
approach, the generic code can manipulate the generic portions of a
tree directly, which allows large chunks of a language-parametric
transformation to be written similarly to a normal single-language rewrite.

Sort injections are an instance of the concept of \emph{feature interactions}  from the field of software product lines \cite{van2001notion}. A similar idea is seen in the TruffleVM \cite{grimmer2015high} to allow language runtimes to exchange messages.

The past decade has seen extensive work in \emph{language workbenches}, which are designed to make it easy to implement languages and transformations on them. They include Spoofax and its component Stratego \cite{kats2010spoofax}, Rascal \cite{klint2009easy}, TXL \cite{cordy2006txl}, Semantic Designs DMS \cite{baxter2004dms}, and JetBrains MPS \cite{voelter2012language}. These were extensively surveyed in Erdweg et al \cite{erdweg2013state}. All these share the limitation that, while they make it easy to define languages and write transformations, the resulting transformations can only run on one representation of one language. At best they can be used to implement the ``Clang-style'' common representation, discussed in Section \ref{sec:why-irs-fail}.

One recent work that echoes our own is Brown et al's
\cite{brown2016build} work  using \emph{island grammars}
\cite{moonen2001generating} to write static analyzers for multiple
languages. They show that they only need to represent fragments of a
language to construct an analyzer. Their analyzers are still built for
a single language, and they resort to cloning code to implement them
for others. They do not address transformation. 

Incremental concrete syntax \cite{dinkelaker2013incremental} is a technique using island grammars to construct parsers. It focuses on concrete syntax (parsing); ours is on abstract syntax (representation).

\section{Conclusion}

Incremental parametric syntax fulfills a simple promise: when writing
similar transformations for multiple languages, they should be able to
share code to the extent the languages are similar. We think that the ability to make multi-language transformation tools will greatly increase the cost/benefit ratio of building tools, and other researchers are noticing. In our previous presentations of \textsc{Cubix}, we were approached by groups from Microsoft Research and Uber who wished to implement our approach to support their own multi-language tooling, while several other groups inquired about using \text{Cubix} itself in their research. One from the University of Washington has already begun doing so. We plan to work extensively on supporting these efforts after the public release of \textsc{Cubix}.

The work in this paper is, to our knowledge, the first to allow a single program to perform source-to-source transformations on multiple real languages while preserving the information of each. We believe incremental parametric syntax solves a key problem in writing multi-language tools. The \textsc{Cubix} framework is available from:

\begin{center}
\url{https://github.com/jkoppel/cubix}
\end{center}

\noindent The RWUS suite is available from:

\begin{center}
\url{https://github.com/jkoppel/rwus}
\end{center}

\begin{acks}

We thank Chris Barnett and Jonathan Paulson for assisting in the construction of the RWUS test suite, and to Dieter Vekeman for helping to fix pretty-printer bugs in our dependencies. We thank Carrie Cai, Jiasi Shen, and Ethan Bian for advising in the experimental and visual design of the study. We further thank our other \textsc{Cubix} contributors, Jasper Haag and M\'at\'e Kov\'acs. Finally, we thank everyone who has given us feedback on earlier drafts of the paper, especially the anonymous reviewers, and the many audiences who have given us feedback on presentations of \textsc{Cubix}.

This material is based upon work supported by the National Science Foundation Graduate Research
Fellowship under Grant No. 1122374.
\end{acks}

\bibliography{compstrat.bib,hackage.bib}

\clearpage

\appendix

\setlength{\belowcaptionskip}{0pt}
\setlength{\floatsep}{12pt}
\setlength{\textfloatsep}{20pt}
\setlength{\intextsep}{12pt}

\section{The ADT Modularization Transformation}
\label{app:adt-modularization}

\FloatBarrier

This section gives a formal definition for the ADT modularization
transformation implemented in \code{comptrans}. This algorithm
transforms a syntax definition given as a family of mutually recursive
ADTs into an equivalent definition in the sum-of-signatures
representations, expressed as a collection of GADT definitions
combined with an explicit sum and fixpoint.

Figure \ref{fig:gadt-syntax} gives a syntax for GADTs.  GADTs and ADTs
are given as a set of constructors with a given type. We assume that
there is a corresponding set of type constructors. We assume that ADTs
have monomorphic types, and their associated constructors have the
base kind. Conversely, GADTs may have polymorphic types with equality constraints, and their
constructors may have higher kinds. We use the syntax $\forall \overline{\nu} :
\overline{k}. \overline{D} \Rightarrow \sigma$ as sugar for nested
forall types. The language includes two ``container'' functors, lists
and pairs, to give an example of how the transformation deals with
containers embedded in syntax trees. Figure \ref{fig:gadt-kinding} gives the kinding rules
for ADTs and GADTs which check if the constructor types are
well-formed. $\Gamma$ is a local context storing all type variables in
scope, while we assume $\Phi$ has been populated with the types of all
constructors declared.

To close a set of GADT definitions into a sum-of-signatures
representation of a syntax, the language must be extended with sum and
recursive types, and with the ability to instantiate a polymorphic
type. Figure \ref{fig:gadt-composition} gives this extension and the
corresponding typing rules. Note that the sum types in this language
are of polymorphic kind. To avoid the need to track constraints with
variables, the rule for polymorphic type application recurses into the
left-hand side until the constraint can be checked syntactically. This
has the unfortunate consequence that kind-checking may become circular. These typing rules should hence be interpreted with greatest fixed point
semantics, meaning that circularly-defined judgments hold.

Figure \ref{fig:trans-algo} gives the transformation algorithm. The
transformation replaces every ADT constructor with a type of kind $*$ with a GADT
constructor with a type constructor of kind $(*\rightarrow *)\rightarrow * \rightarrow
*$. More importantly, these GADT type constructors do not refer to any
other type definitions, with the exception of the ``label'' types,
which are purely nominal and uninhabited. It makes use of three
auxiliary functions. $\text{newcon}(\text{con})$ returns a fresh name for a GADT
constructor. $\text{newconType}(\nu)$ does similar for the
corresponding type constructors. $\text{lab}(C)$ maps each constructor $C$ to a
corresponding ``label'' constructor of kind $*$. 

Of final note, in order to inhabit terms of sort $\text{List}\:
\gamma$ or $\text{Pair}\:\gamma\:\iota$, there are three
specially-defined constructors, given in Figure
\ref{fig:gadt-container-functors}. These are given Curry-style types,
meaning their types contain an unbound variable, and they may be given
a type for any instantiation of that variable.

We are now ready to state the property this transformation was
designed to satisfy: For a family of mutually recursive ADTs
defined by $\overline{con^\sigma}$ with root type $C$, $C$ is
equivalent to the sum of the generated GADTs at sort
$\text{lab}(C)$. Formally, if $\Phi$ contains the types for all
declared ADT constructors and
$\Phi\entails\overline{con^\sigma}\text{ okay}$, then $$\Phi \entails C \equiv  (\mu \alpha : *\rightarrow *
. (\text{PairF}+\text{ListF}+\sum_{s\in \overline{\sigma}}
\text{transType}(s))\:\alpha ) \: \text{lab}(C)$$

Here, $\equiv$ denotes the classical notion of a type isomorphism, i.e.: the presence of a pair of mutually inverse functions that convert from one to the other. This is still an informal statement, albeit in formal notation, as we have not fully defined the language of terms which is needed to make this statement fully rigorous. The description in this section is meant only to unambiguously describe the algorithm.


\begin{figure}
\begin{mdframed}
\begin{center}
\begin{tabular}{rl}
Type variables & $\alpha, \beta, \dots$ \\
Predefined constructors & C \\
\\
Kinds & $k \sdef * \sor k \rightarrow k$ \\
& \\
Primitive types & $P \sdef \text{Int} \sor \text{Bool} \sor
\dots $ \\
Container functors & $F \sdef \text{List} \sor \text{Pair}$ \\
\\
Base types & $\nu \sdef  \alpha \sor C \sor F \sor P \sor \nu \nu$ \\
Monotypes & $\tau \sdef \nu \sor \nu \rightarrow \tau$ \\
Constraints & $D \sdef \cdot \sor \alpha \sim \nu$ \\
Polytypes & $\sigma \sdef \tau \sor \forall \alpha :
k . D \Rightarrow \sigma$ \\
& \\

Constructors & $c \sdef \text{con}^\sigma$
\end{tabular}
\end{center}
\end{mdframed}
\caption{A syntax for GADTs}
\label{fig:gadt-syntax}
\end{figure}

\begin{figure*}
\begin{mdframed}
\begin{center}
\begin{tabular}{l}
\infer[PRIM]
  {\Gamma; \Phi\entails P : *}
  {}  \\
 \\
\vspace{1em}
\infer[LIST]
  {\Gamma; \Phi\entails\text{List} : *\rightarrow *}
  {} \\
\vspace{1em}
\infer[PAIR]
  {\Gamma;\Phi\entails\text{Pair} : * \rightarrow * \rightarrow *}
  {} \\
\vspace{1em}
\infer[VAR]
  {\Gamma;\Phi\entails \alpha : k}  
  {\alpha : k \in \Gamma} \\
\vspace{1em}
\infer[CON]
  {\Gamma; \Phi\entails C : k}
  {C : k \in \Phi} \\
\vspace{1em}
\infer[ARR]
  {\Gamma; \Phi \entails \nu \rightarrow \tau : *}
  {\Gamma;\Phi \entails \nu : * & \Gamma; \Phi \entails \tau : *} \\
\vspace{1em}
\infer[FORALL]
  {\Gamma; \Phi \entails \forall \alpha : k_1. D\Rightarrow \sigma : k_1\rightarrow k_2}
  {\Gamma, \alpha : k_1 ; \Phi \entails \sigma : k_2} \\
\vspace{1em}    
\infer[APP]
  {\Gamma; \Phi \entails \nu_1\nu_2 : k_2}
  {\Gamma; \Phi \entails \nu_1 : k_1 \rightarrow k_2 & \Gamma; \Phi
    \entails \nu_2 : k_1} \\
\vspace{1em}

\infer
  {\Phi \entails \isok{\text{con}^\sigma}}
  {\emptycxt; \Phi \entails \sigma : *} \\

\end{tabular}
\end{center}
\end{mdframed}
\caption{}
\label{fig:gadt-kinding}
\end{figure*}

\begin{figure*}
\begin{mdframed}

\begin{center}
$\sigma \sdef \dots \sor \sigma + \sigma \sor \mu \alpha:k.\; \sigma \sor  \sigma\sigma$
\\[2em]

\begin{tabular}{l}
\infer[POLY-SUM]
  {\Gamma; \Phi \entails \sigma_1 + \sigma_2 : k}
  {\Gamma; \Phi \entails \sigma_1 : k & \Gamma; \Phi \entails
    \sigma_2 : k} \\
 \\
\vspace{1em}
\infer[REC]
  {\Gamma; \Phi \entails \mu\alpha : k. \sigma : k}
  {\Gamma, \alpha : k; \Phi \entails \sigma : k} \\
\vspace{1em}
\infer[POLY-APP-SUM-LEFT]
  {\Gamma,\Phi \entails (\sigma_1 + \sigma_2)\sigma_3 : k}
  {\Gamma, \Phi \entails \sigma_1 + \sigma_2 : k^\prime & \Gamma,\Phi \entails \sigma_1 \sigma_3 : k} \\
\vspace{1em}
\infer[POLY-APP-SUM-RIGHT]
  {\Gamma,\Phi \entails (\sigma_1 + \sigma_2)\sigma_3 : k}
  {\Gamma, \Phi \entails \sigma_1 + \sigma_2 : k^\prime & \Gamma,\Phi \entails \sigma_2 \sigma_3 : k} \\
\vspace{1em}

\infer[POLY-APP-REC]
  {\Gamma,\Phi \entails (\mu\alpha : k.\;\sigma)\sigma^\prime : k^\prime}
  {\Gamma, \Phi \entails (\mu\alpha : k. \sigma) : k & \Gamma,\Phi \entails (\sigma[(\mu\alpha : k.\;\sigma) / \alpha]) \sigma^\prime
                                        : k^\prime} \\
\vspace{1em}
\infer[POLY-APP]
  {\Gamma,\Phi \entails (\forall \alpha :
    k^\prime. \cdot\Rightarrow \sigma) \nu : k}
  {\Gamma,\Phi\entails \nu : k^\prime & \Gamma,\Phi \entails
                                        \sigma [\nu /\alpha ] :
                                        k} \\
\vspace{1em}
\infer[POLY-APP-CONSTRAINT]
  {\Gamma,\Phi \entails (\forall \alpha :
    k^\prime. \alpha\sim \nu\Rightarrow \sigma) \nu : k}
  {\Gamma,\Phi\entails \nu : k^\prime & \Gamma,\Phi \entails \sigma [\nu /\alpha ] : k} \\
\end{tabular}
\end{center}
\end{mdframed}
\caption{}
\label{fig:gadt-composition}
\end{figure*}

\begin{figure*}
\begin{mdframed}
\begin{center}
\vspace{2em}
\begin{align*}
\text{ConsF} &:&& \forall \: (\alpha : *\rightarrow *)\:(\gamma :
*). \gamma \sim List \:\iota \Rightarrow \alpha\:\iota\:
\rightarrow\: \alpha\: \gamma\:\rightarrow
\text{ListF}\:\alpha\: \gamma &&\text{ for any }\iota\\
\text{NilF} &:&& \forall \: (\alpha : *\rightarrow *)\:(\gamma :
*). \gamma \sim List \:\iota \Rightarrow \text{ListF}\:\alpha\: \gamma&&\text{ for any }\iota\\
\text{PairF} &:&& \forall (\alpha : *\rightarrow *)\:(\gamma :
*). \gamma \sim (\text{Pair } \iota\:\kappa) \Rightarrow
\alpha \: \iota \rightarrow\:\alpha\:\kappa\rightarrow
\text{PairF}\:\alpha\: \gamma&&\text{ for any }\iota,\kappa \\
\end{align*}
\end{center}
\end{mdframed}
\caption{}
\label{fig:gadt-container-functors}
\end{figure*}

\begin{figure*}
\begin{mdframed}
\begin{center}
\vspace{2em}
\fbox{$\text{trans}(\text{con}^\tau)$} \\
$\text{trans}(\text{con}^\tau)
=\text{newcon}(\text{con})^{\text{transTypeTop}(\tau)}$ \\
\vspace{2em}
\fbox{$\text{transTypeTop}(\tau)$} \\
$\text{transTypeTop}(\tau) = \forall (\alpha : *\rightarrow *).\;\cdot\; \Rightarrow (\forall \:(\gamma : *). \;\gamma\sim\text{getSort}(\tau) \; \Rightarrow\text{transType}(\tau,\alpha,\gamma))$
\vspace{2em}

\fbox{$\text{getSort}(\tau)$} \\
\begin{tabular}{ll}
$\text{getSort}(\nu\rightarrow\tau) $&$= \text{getSort}(\tau)$ \\
$\text{getSort}(C) $&$= \text{lab}(C)$ \\
\end{tabular} \\
\vspace{2em}

\fbox{$\text{transType}(\tau,\alpha,\gamma)$} \\
\begin{tabular}{ll}
$\text{transType}(\nu\rightarrow\tau,\alpha,\gamma) $&$=
\text{transTypeBase}(\nu,\alpha,\gamma)\rightarrow\text{transType}(\tau,\alpha,\gamma)$
\\
$\text{transType}(C,\alpha,\gamma) $&$=
  \text{newconType}(C)\:\alpha\:\gamma$\\
\end{tabular} \\
\vspace{2em}
\fbox{$\text{transTypeBase}(\nu,\alpha,\gamma)$} \\
\begin{tabular}{ll}
$\text{transTypeBase}(P,\alpha,\gamma) $&$= P$ \\
$\text{transTypeBase}(F\overline{\nu},\alpha,\gamma) $&$= \alpha\:(F\:\overline{\text{transTypeBase}(\nu)})$ \\
$\text{transTypeBase}(C,\alpha,\gamma) $&$= \alpha\:\text{lab}(C)$ \\
\end{tabular}
\end{center}
\end{mdframed}
\caption{}
\label{fig:trans-algo}
\end{figure*}

\FloatBarrier

\section{Readability Study: Full Details}
\label{app:study-long}

We ran a study to evaluate the readability of our transformations' output. The overall setup of our experiment is like a Turing test. First, we
ask a set of human contributors to transform programs by hand. We then
give a separate set of human judges these programs, alongside the
corresponding automatically transformed programs, and ask them to rate them both on correctness and quality. Because low-level code formatting is outside the scope of our claims, we automatically reformat the human-written code before presenting them for comparison. Outside of formatting, we attempted to bias the experiment in favor of the humans, allowing them to resubmit until their transformed programs were correct according to our extremely thorough test suites. Despite this, in our final results, the judges gave the automatically transformed programs a higher average rating.

Our experiment proceeds in three phases. In the first phase, we construct the RWUS suite, providing suitable programs on which to run the study. In the second phase, we ask human participants to manually apply each of the three studied transformations on a code sample. In the final phase, human judges from Mechanical Turk rate the manually-transformed code against the same code transformed by our system. Note that this study was completed using earlier versions of the transformations which failed some tests.

\subsection{Phase 1: Constructing the RWUS Suite}

As objects in our study, we needed (1) representative samples of real-world code, and (2) an objective measure of whether the code was transformed correctly. The second criterion is the main difficulty, as random samples of code typically do not come with thorough tests, and certainly not tests that are easy to run. Hence, we created our own.

The RWUS (Real World, Unchanged Semantics) suite consists of 50 functions across 5 languages randomly selected from top GitHub projects. For each, it also includes a test suite designed with the intention that only functions semantically equivalent to that function will pass. Each function is distributed as an {\it entry}. An entry is a file containing the original sample, mocks for all referenced symbols, tests, and a wrapper \code{main} procedure which invokes
the tests. The files can all be compiled and executed without any dependencies. The tests are used by invoking a script that replaces the sample with a transformed version, and then executes the resulting file.

We selected the functions for the RWUS suite as follows: For each of C, Java, JavaScript, Lua, and Python, we downloaded the top 20 projects in that language on GitHub from those with at least 500 lines, sorted by number of users who ``starred'' that project. We then uniformly at random selected a line of code from the projects. If this line of code lies within a function, we took the innermost such function as a sample; else, we repeated the process. We discarded all samples which were not between 5 and 50 lines of code, excluding function signatures, blank lines, and comments. We repeated this process until we had $10$ samples for each language. One shortcoming of this approach is that the top-rated projects on GitHub vary in size by orders of
magnitude. As the extreme, 90\% of our C corpus and all 10 C samples come from Linux. The other 40 samples come from 24 different projects.

For each sample, we constructed test cases ensuring full path coverage, and added checks to ensure all mocked functions are called in the expected order with the expected arguments. The resulting tests are incredibly thorough. While the actual samples total 1158 lines of code, the RWUS suite totals 8070 lines of code.

The RWUS suite is available from:

\begin{center}
\url{https://github.com/jkoppel/rwus}
\end{center}


\subsection{Phase 2: Obtaining Human-Written Transformations}

We recruited programmers through department mailing lists, flyers
posted around the department, and social media. Due to the relative scarcity of Lua programmers, we also
posted on Lua forums, and asked Lua participants to spread the study by word of mouth.

Participants were sent to a website, where they would download a
single sample from the RWUS suite along with its tests, and were asked to perform each of
our transformations by hand on the file. They were allowed to contribute one sample per language, and were offered a \$10
Amazon gift card for each.

We inspected each submission by hand. Participants were asked to
resubmit until their transformed samples passed all tests, and had no
significant transformation errors, such as unhoisted variables. 

\subsection{Preparing the Samples}

After we had collected all 50 human-transformed samples, we ran them through the corresponding parser and pretty printer to normalize formatting. We then ran
our transformations on each of the RWUS samples, and evaluated them with the RWUS test suites.

We did not run any transformation on
the RWUS samples until all development on the transformations had ceased. We also attempted to avoid allowing knowledge of the samples in
the RWUS suite to influence development of the transformations,
although a single researcher was responsible for both.

Of the 120 transformed pairs, for 24 of them, the automatically transformed version was identical to the human written one after reformatting. These are broken down per language in Table \ref{table:transform-same}. Six automatically transformed samples either failed their test suites or caused an error in the transformed program, while, for one sample, a pretty-printer bug caused both the human-transformed and automatically transformed versions to fail to compile. The remaining 89 pairs were sent to human judges for evaluation in Phase 3.

\begin{table}
\begin{center}
\caption{Counts of programs where presentation to the human judges was
  inappropriate}
\label{table:transform-same}
\begin{tabular}{r|rrrrr}
\hline
& C & Java & JS & Lua & Python \\
\hline
{\bf Identical} & 6 & 9 & 1 & 5 & 3 \\
{\bf Failed} & 0 & 1 & 4 & 0 & 1 \\
\hline
\end{tabular}
\end{center}
\vspace{2pt}
\end{table}

\subsection{Phase 3: Comparing Human and Machine-Written Transformations}

In Phase 3, we asked human judges from Mechanical Turk to rate the manually-transformed code from Phase 2 along with their automatically transformed counterparts.

We created one task on Mechanical Turk for each
language/transformation combination. For each judge entering our website interface, we began by
presenting an explanation and example of the transformation, before presenting the questions. Each question shows a sample program, along with the automatically transformed version produced by our system, and the manually-transformed version collected in Phase 1. They were asked to rate both on a 1--5 scale. We instructed that
they should first rate the transformed programs on correctness vs. the
original program, second on faithfulness to the intended transformation, and only third on general prettiness and code quality.  Both the order of questions and the order of the transformed pairs were randomized. We assigned each of the 20-30 transformed samples to 10 judges, giving us up to 300 ratings per language.

\begin{figure*}
\begin{center}
\includegraphics[scale=0.4]{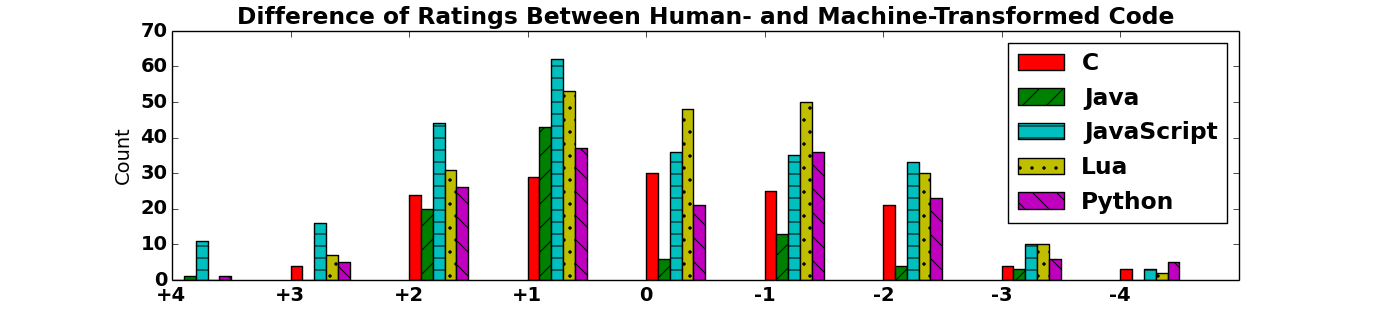}
\end{center}
\caption{Counts of differences between the ratings of the machine transformations and the human transformations. The leftmost bars represent cases where the judge rated the machine-produced output higher than the human-produced.}
\label{fig:human-study-results-appendix}
\end{figure*}

\subsection{Quality Control}

The setup described above does not preclude someone from rating programs randomly, so we employed two quality-control mechanisms. Our primary form of quality control was the creation of ``canary'' questions. Canaries appear as normal questions, except that the programs contained therein were contrived. In two of the canaries, one of the programs was clearly not a transformed version of the original. In the third canary, both displayed programs were identical. We rejected any submission in which the worker did not rate the correct program higher for the first two canaries, or did not rate both programs of the third canary the same. Second, if a worker ever submitted two answers within 11 seconds of each other, we marked this worker as untrustworthy, and rejected all submissions by him. We picked this value after observing the times spent on each question in dry runs of the study.

We noticed substantial differences between workers who did and did not pass the quality controls. Workers with one rejected submission typically had rejected submissions for many different languages. Workers with accepted submissions were much more likely to only submit for one language. Workers typically either had all their submissions accepted or all rejected. Furthermore, we noticed that rejected submissions were typically completed in much less time than accepted ones, although many workers who failed the canaries were substantially slower than the fastest correct workers.

The experimenters manually inspected a selection of judgments from accepted submissions, and found them all reasonable. Overall, our observations suggest that our quality control mechanisms did effectively classify workers on skill, and that our data is high-quality.

\subsection{Results}

For each language, we tabulated the difference in ratings between the human-written and automatically transformed programs. Our results are given in Figure \ref{fig:human-study-results-appendix}. The average differences in ratings ranged from $-0.075$ for Python (favoring the humans) to $+0.633$ for Java (favoring the machine). The differences for C, JavaScript, and Lua were $-0.014$, $+0.396$, and $-0.052$ respectively.

Our goal was to show that the output of our transformations is not less readable than the human-transformed code. This is a problem in statistics known as non-inferiority testing \cite{wellek2010testing}. For each language, we formulated a hypothesis that the average difference in ratings between each the machine- and human-transformed code is at least $-1$. We then factored in the pairs that were not sent to Phase 3: each identical pair was counted as 10 judgments of equality (difference $0$), and each pair where the machine-transformed version was incorrect was counted as 10 judgments that maximally penalize the machine version (difference $-4$). We tested each of the $5$ hypotheses using a paired t-test. For each language, it showed that the machine-transformed code was non-inferior by a non-inferiority margin of at most $1$ with $p<10^{-8}$. In retrospect, this data had the power to prove the hypothesis with a much smaller non-inferiority margin.

Considering both the raw data and the statistical tests, our study provides strong evidence that the output of transformations in \textsc{Cubix} is no less readable than hand-transformed code, showing that implementing source-to-source transformations with incremental parametric syntax avoids the mangling common to IR-based approaches.

\subsection{Threats to Validity}

Our results are potentially biased by using a real-world distribution of programming constructs, as opposed to intentionally constructing a suite filled with corner cases. The humans are hindered by a lack of learning: they only perform each transformation once per language. Finally, we cannot be certain of the quality of the data from Mechanical Turk. In our dry runs, we found that workers on Mechanical Turk tend to rate simple programs more highly, even when the transformation is incorrect. Two of our canaries are specifically designed to prevent this behavior.

\end{document}